\documentclass[aps,pra,twocolumn,superscriptaddress,nobibnotes,tightenlines,longbibliography]{revtex4-1}
\usepackage[utf8]{inputenc}
\usepackage[T1]{fontenc}
\usepackage{amsmath}
\usepackage{amssymb}
\usepackage{amsfonts}
\usepackage{mathrsfs}
\usepackage{bm, dsfont}
\usepackage{booktabs}
\usepackage{multirow}
\usepackage[usenames,dvipsnames]{color}
\usepackage{colortbl}
\usepackage[table]{xcolor}
\usepackage{enumitem}
\usepackage{dcolumn}
\usepackage{graphicx}
\usepackage{natbib}
\usepackage[colorlinks=true,linkcolor=blue,citecolor=blue,urlcolor=blue]{hyperref}
\usepackage{hypcap}
\usepackage{verbatim}
\usepackage{psfrag}
\usepackage[normalem]{ulem}
\usepackage{siunitx}

\graphicspath{{./}}

\newcommand{\bra}[1]{{\langle{#1}|}}
\newcommand{\ket}[1]{{|{#1}\rangle}}
\newcommand{\ketbra}[2]{{|{#1}\rangle\!\langle{#2}|}}
\newcommand{\braket}[2]{{\langle{#1}|{#2}\rangle}}
\newcommand{\ev}[1]{\langle{#1}\rangle}
\newcommand{\tr}{\text{tr}}

\newcolumntype{C}{>{$}c<{$}}
\AtBeginDocument{
\heavyrulewidth=.08em
\lightrulewidth=.05em
\cmidrulewidth=.03em
\belowrulesep=.65ex
\belowbottomsep=0pt
\aboverulesep=.4ex
\abovetopsep=0pt
\cmidrulesep=\doublerulesep
\cmidrulekern=.5em
\defaultaddspace=.5em
}

\begin{document}
\title{Heralded distribution of single-photon path entanglement}

\author{P.~Caspar}
\thanks{These authors contributed equally to this work.}
\affiliation{Department of Applied Physics, University of Geneva, CH-1211 Genève, Switzerland}
\author{E.~Verbanis}
\thanks{These authors contributed equally to this work.}
\affiliation{Department of Applied Physics, University of Geneva, CH-1211 Genève, Switzerland}
\author{E.~Oudot}
\affiliation{Department of Applied Physics, University of Geneva, CH-1211 Genève, Switzerland}
\affiliation{Quantum Optics Theory Group, University of Basel, CH-4056 Basel, Switzerland}
\author{N.~Maring}
\affiliation{Department of Applied Physics, University of Geneva, CH-1211 Genève, Switzerland}
\author{F.~Samara}
\affiliation{Department of Applied Physics, University of Geneva, CH-1211 Genève, Switzerland}
\author{M.~Caloz}
\affiliation{Department of Applied Physics, University of Geneva, CH-1211 Genève, Switzerland}
\author{M.~Perrenoud}
\affiliation{Department of Applied Physics, University of Geneva, CH-1211 Genève, Switzerland}
\author{P.~Sekatski}
\affiliation{Quantum Optics Theory Group, University of Basel, CH-4056 Basel, Switzerland}
\author{A.~Martin}
\altaffiliation[Present address: ]{Université Côte d'Azur, CNRS, Institut de Physique de Nice, Parc Valrose, F-06108 Nice Cedex 2, France}
\affiliation{Department of Applied Physics, University of Geneva, CH-1211 Genève, Switzerland}
\author{N.~Sangouard}
\affiliation{Quantum Optics Theory Group, University of Basel, CH-4056 Basel, Switzerland}
\affiliation{Institut de physique théorique, Université Paris Saclay, CEA, CNRS, F-91191 Gif-sur-Yvette, France}
\author{H.~Zbinden}
\affiliation{Department of Applied Physics, University of Geneva, CH-1211 Genève, Switzerland}
\author{R.~T.~Thew}
\email[Electronic address: ]{Robert.Thew@unige.ch}
\affiliation{Department of Applied Physics, University of Geneva, CH-1211 Genève, Switzerland}
\begin{abstract}
We report the experimental realization of heralded distribution of single-photon path entanglement at telecommunication wavelengths in a repeater-like architecture. The entanglement is established upon detection of a single photon, originating from one of two spontaneous parametric down-conversion photon pair sources, after erasing the photon's which-path information. In order to certify the entanglement, we use an entanglement witness which does not rely on postselection. We herald entanglement between two locations, separated by a total distance of \SI{2}{km} of optical fiber, at a rate of \SI{1.6}{kHz}. This work paves the way towards high-rate and practical quantum repeater architectures.
\end{abstract}
\maketitle
Sharing photonic entanglement over long distances is a key resource for building a quantum communication network~\cite{Kimble2008,Wehner2018}. In order to distribute entanglement to two remote parties through optical fiber, quantum repeater schemes provide a solution to overcome the direct transmission loss~\cite{Briegel1998}. The basic idea is to divide the whole distance into elementary links in each of which entanglement is independently established in a heralded way between two quantum memories. Finally, successive entanglement swapping operations between the links are used to extend the entanglement over the whole distance.
Among the different quantum repeater schemes, those using single-photon path entanglement~\cite{Tan1991}, where a single photon is delocalized into two modes, are promising candidates for establishing such a network since they require fewer resources as well as being less sensitive to memory and detector inefficiencies compared to other repeater schemes due to their linear, rather than quadratic, loss scaling~\cite{Sangouard2011}. \par 
A proposed postselection-free approach for entanglement distribution, based on the erasure of the heralding photon's which-path information, is a modification of the Duan-Lukin-Cirac-Zoller (DLCZ) protocol~\cite{Duan2001} that employs photon pair sources and multimode memories~\cite{Simon2007}. A practical implementation of this scheme, however, faces two major challenges. It requires first, stabilization and control of the optical phase between the two parties, and second, a practical implementation for entanglement certification. 
Experiments overcoming both challenges by employing individual matter qubits have been presented with ions~\cite{Slodicka2013}, quantum dots~\cite{Delteil2016,Stockill2017}, and nitrogen-vacancy centers~\cite{Humphreys2018}. 
However, in all those table-top demonstrations, entanglement was heralded by a photon outside the telecom band and thus will need to be frequency converted, which will further complicate the management of phase in the network. 
An approach combining an atomic ensemble quantum memory at near-infrared wavelengths and quantum frequency conversion to the telecom O-band has recently been reported~\cite{Yu2020}, however, the entanglement was certified by recombining the entangled modes (single-photon interference) which is not applicable in a distributed scenario. \par
\begin{figure}
\capstart
\includegraphics[width = 8.6cm]{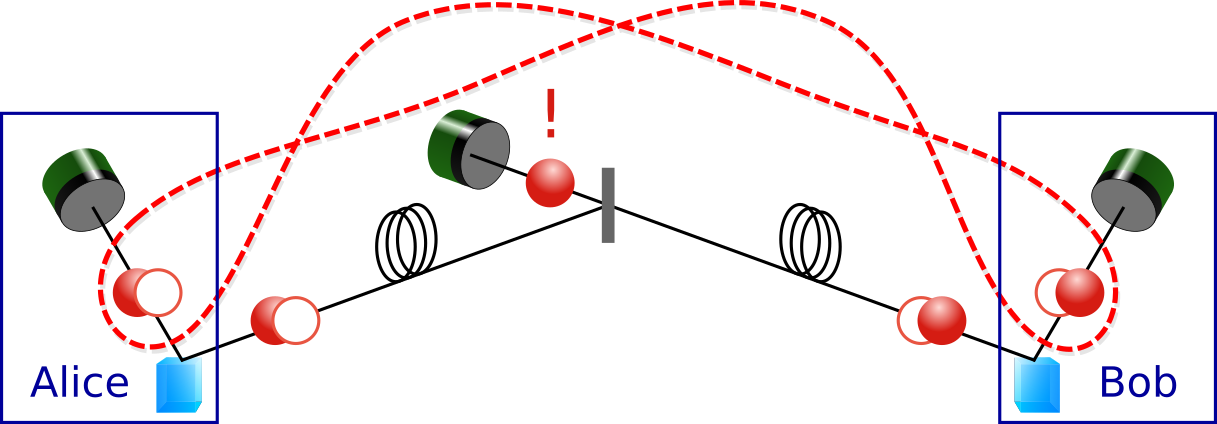}
\caption{\label{fig:Concept} Conceptual schematic of the experiment. A successful detection of a photon at the central station, originating from one of two photon pair sources, heralds the distribution of a single-photon path-entangled state between Alice and Bob.}
\end{figure}

 
In this work we demonstrate a scheme for the heralded distribution of single-photon path entanglement at telecom wavelengths over a distance of $2\!\times\!\SI{1.0}{km}$ of optical fiber, in a quantum repeater-like architecture (see Fig.~\ref{fig:Concept}). The detection of a single photon at the central station erases the which-path information about which one of the two photon pair sources it was emitted from, and heralds the distributed entangled state. The fiber connecting Alice and Bob is part of a phase stabilized interferometer. Another fiber (not shown), between Alice and Bob, closes the interferometer and is connected to a laser that is used to stabilize the interferometer. Inspired by an entanglement witness~\cite{Monteiro2015} using displacement-based photon detections~\cite{Vivoli2015}, the distributed entanglement is measured locally and certified by an entanglement witness that is robust to loss, and does not make assumptions about the state itself. \par
\emph{Concept.}---Each of the two spontaneous parametric down-conversion (SPDC) photon pair sources, held by Alice and Bob, creates a two-mode squeezed vacuum state with low photon pair creation probability per pump pulse $P_\mathrm{pair,A}=P_\mathrm{pair,B} \ll 1$. Two modes, one from each source, are combined on a 50/50 beam splitter at the central station. By neglecting contributions from higher order pair creation probabilities $\mathcal{O}(P_\mathrm{pair})$, the resulting state shared between Alice and Bob, conditioned on the detection of a heralding photon, can be written as 
\begin{equation}\label{eq:state}
\ket{\psi}_{AB} = \frac{1}{\sqrt{2}}\left(\ket{10}_{AB}+e^{i(\theta_{B}-\theta_{A})}\ket{01}_{AB}\right),
\end{equation}
where $\ket{0}$ denotes the vacuum state, $\ket{1}$ the single-photon number state, $\theta_{A(B)}=\phi_{A(B)}+\chi_{A(B)}$ with $\phi_{A(B)}$ the phase of the pump before the source on Alice's (Bob's) side and $\chi_{A(B)}$ the respective phase acquired by the photon traveling from the source to the central station [see Ref.~\cite{Simon2007} for the derivation of Eq.~\eqref{eq:state} and a more complete discussion]. The first conceptual challenge of this scheme is to prove entanglement within the $\{\ket{0},\ket{1}\}$ subspace. This is possible using photon detection techniques preceded by weak displacement operations~\cite{Paris1996,Banaszek1999,Kuzmich2000,Bjork2001,Hessmo2004,Vivoli2015}, as described in the following. \par

\emph{Displacement-based measurement.}---We introduce the bosonic annihilation and creation operators $a_i$ and $a_i^{\dagger}$ with $i \in \{1,2\}$ acting on the photonic modes on Alice's ($i=1$) and Bob's ($i=2$) sides. We assume non-photon-number-resolving detectors, that is, only two different measurement results can be produced in each run. A ``no-click'' event is modeled by a projection on the vacuum state $\ketbra{0}{0}$ whereas a ``click'' event corresponds to the projection into the orthogonal subspace $\mathds{1}-\ketbra{0}{0}$. If we attribute the outcome $+1$ to a no-detection and $-1$ to a conclusive detection, the observable including the displacement operation $D(\alpha_i)=e^{\alpha_i a_i^{\dagger}-\alpha_i^\ast a_i}$ on mode $i$ is given by
\begin{equation}\label{eq:sigma}
 \sigma_{\alpha_i}^{(i)}=D^{\dagger}(\alpha_i)
 (2\ketbra{0}{0}-\mathds{1})D(\alpha_i).
\end{equation}
In the qubit subspace $\{\ket{0},\ket{1}\},$ $\sigma_{0}^{(i)}$ corresponds exactly to the Pauli matrix $\sigma_z$ on mode $i$. When $\alpha$ increases in amplitude, the positive operator valued measure (POVM) elements associated with outcomes $+1$ get closer to projections in the $x-y$ plane of the Bloch sphere~\cite{Vivoli2015}. For $\alpha=1$ ($\alpha=i$), these POVM elements are projections along non-unit vectors pointing in the $x \,(y)$ direction. \par 

\emph{Entanglement certification.}---Building upon the displacement-based measurement, we now elaborate on the theory behind the witness that we developed to optimally certify path entanglement even in a lossy environment. The certification of entanglement for the state $\ket{\psi}_{AB}$ requires access to the coherence terms $\ketbra{01}{10}$ and $\ketbra{10}{01}$ as well as to the probabilities $P_{ij}$ to have $i$ photons on the first mode and $j$ photons on the second mode. A good entanglement witness for our state is the observable
\begin{equation}\label{eq:W}
\hat{\mathcal{W}} = \sigma_{\alpha_1}^{(1)}\otimes\sigma_{\alpha_2}^{(2)}, 
\end{equation}
which is phase averaged according to
\begin{equation}
\label{eq:W_avg}
\hat{W} = \frac{1}{2\pi}\int_0^{2\pi} d\phi\bigg(\prod_{i=1}^2 e^{i\phi\hat{a}_i^\dagger\hat{a}_i}\bigg) \hat{\mathcal{W}} \bigg(\prod_{i=1}^2 e^{-i\phi\hat{a}_i^\dagger\hat{a}_i}\bigg)
\end{equation}
to take into account that from experimental run to run, the global phase of the displacement parameter is arbitrary. Thus, the only remaining coherence terms in $\hat{W}$ are exactly the desired ones between $\ket{01}$ and $\ket{10}$. \par

In order to demonstrate entanglement we use the Peres-Horodecki criterion~\cite{Horodecki1996} stating that separable two-qubit states have a positive partial transpose (PPT). Considering the observable $\hat{W},$ we first compute its maximum expectation value $w_\mathrm{ppt}$ (see Supplemental Material Sec.~II) for separable two-qubit states $\rho_{\text{qubit}}$, i.e. $\rho_{\text{qubit}}\geq 0$ and $\tr(\rho_{\text{qubit}})=1$:
\begin{align}\label{eq:wppt}
w_\mathrm{ppt} &= \max_{\rho_{\text{qubit}}} \tr(\rho_{\text{qubit}}\hat{W}) \\ 
 \text{s.t.} &\quad (i)\; \rho^{T_1}_{\text{qubit}}\geq 0, \nonumber \\
 &\quad (ii)\, \rho_{i,i}=P_{ii}, \nonumber
\end{align}
where $\rho_{i,i}$ denote the diagonal elements of $\rho_{\text{qubit}}$. The advantage of using this witness together with condition $(ii)$ rather than a fixed linear combination of observables as done in Ref.~\cite{Monteiro2015} is that we use all our knowledge of the diagonal elements of the density matrix. This is equivalent to considering a witness constructed from all possible linear combinations of $\sigma_{\alpha_1}^{(1)}\otimes\sigma_{\alpha_2}^{(2)}$, $\sigma_{0}^{(1)}\otimes\sigma_{0}^{(2)}$, $\sigma_{0}^{(1)}\otimes\mathds{1}$ and $\mathds{1}\otimes\sigma_{0}^{(2)}$ which can detect more entangled states. 
If we restrict ourselves to qubit states, the quantities $P_{ij}$ are given by the measured quantities $P_\mathrm{nc,nc}$, $P_\mathrm{nc,c}$, $P_\mathrm{c,nc}$ and $P_\mathrm{c,c}$ without displacement fields where, for example, $P_\mathrm{nc,c}$ is the joint probability of having a ``no-click'' event on the detector on mode 1 and a ``click'' event on mode 2. If the measured value of $\ev{\hat{W}}$ is larger than $w_\mathrm{ppt}$, we can conclude that our state is entangled. \par

To elaborate on the robustness of our witness, let us consider the state $\rho_{\eta}=(1-\eta)\ketbra{00}{00}_{AB}+\eta \ketbra{\psi}{\psi}_{AB}$ which corresponds to adding losses on the state $\ket{\psi}_{AB}$. One can easily check that there always exist settings $\alpha_1$ and $\alpha_2$ such that $\tr\big(\rho_{\eta}\hat{W}\big)>w_\mathrm{ppt}$ for all efficiencies $\eta$ different than 0 (see Supplemental Material Sec.~II). This means that $\hat{W}$ has the ability to detect entanglement for arbitrary loss on the state. Note the fact that we considered detectors with unit efficiencies is still a valid description of our measurement apparatus since one can transfer the inefficiency of the detector to loss on the state (see Supplemental Material Sec.~IV). \par
 
In practice, the amplitude of displacement operations may vary from run to run which could lead to false witness violations. To take into account those fluctuations, we first bound them experimentally and then maximize $w_\mathrm{ppt}$ accordingly, which leads to $\widetilde{w}_\mathrm{ppt}\geq w_\mathrm{ppt}$ (see Supplemental Material Sec.~III). Note that in order to reduce the impact of amplitude fluctuations we choose the mean amplitudes such that $\partial^2 w_\mathrm{ppt}/\partial\alpha_1\partial\alpha_2=0$, which in our case holds true for $\alpha_1=\alpha_2 \approx 0.83$. \par 

So far, we derived the maximum expectation value $w_\mathrm{ppt}$ a separable two-qubit state can achieve. To remove the assumption on the dimension and hence to obtain a state-independent entanglement witness, we derive a general bound for all separable states (see Supplemental Material Sec.~III)
\begin{align}\label{eq:wpptmax}
\begin{split}
w_\mathrm{ppt}^\mathrm{max} &= \widetilde{w}_\mathrm{ppt}+ p_1^\ast+p_2^\ast \\
&\quad +2\beta\sqrt{(p_1^\ast+p_2^\ast)(1-p_1^\ast-p_2^\ast)}
\end{split}
\end{align} 
where $\beta=2 \alpha_1 \alpha_2 e^{-(\alpha_1^2 +\alpha_2^2)}\sqrt{2 (\alpha_1^4 + \alpha_2^4)}$ with $\alpha_i\in \mathbb{R^+}$ and $p_i^\ast$ denote upper bounds on the probabilities of having more than one photon in mode $i$. The latter can be bounded in practice by measuring twofold coincidences after a 50/50 beam splitter. We can thus conclude about entanglement in an arbitrary state $\rho$ if $\tr\big(\rho \hat{W}\big)>w_\mathrm{ppt}^\mathrm{max}$. \par
\begin{figure}
\capstart
\includegraphics[width = 8.6cm]{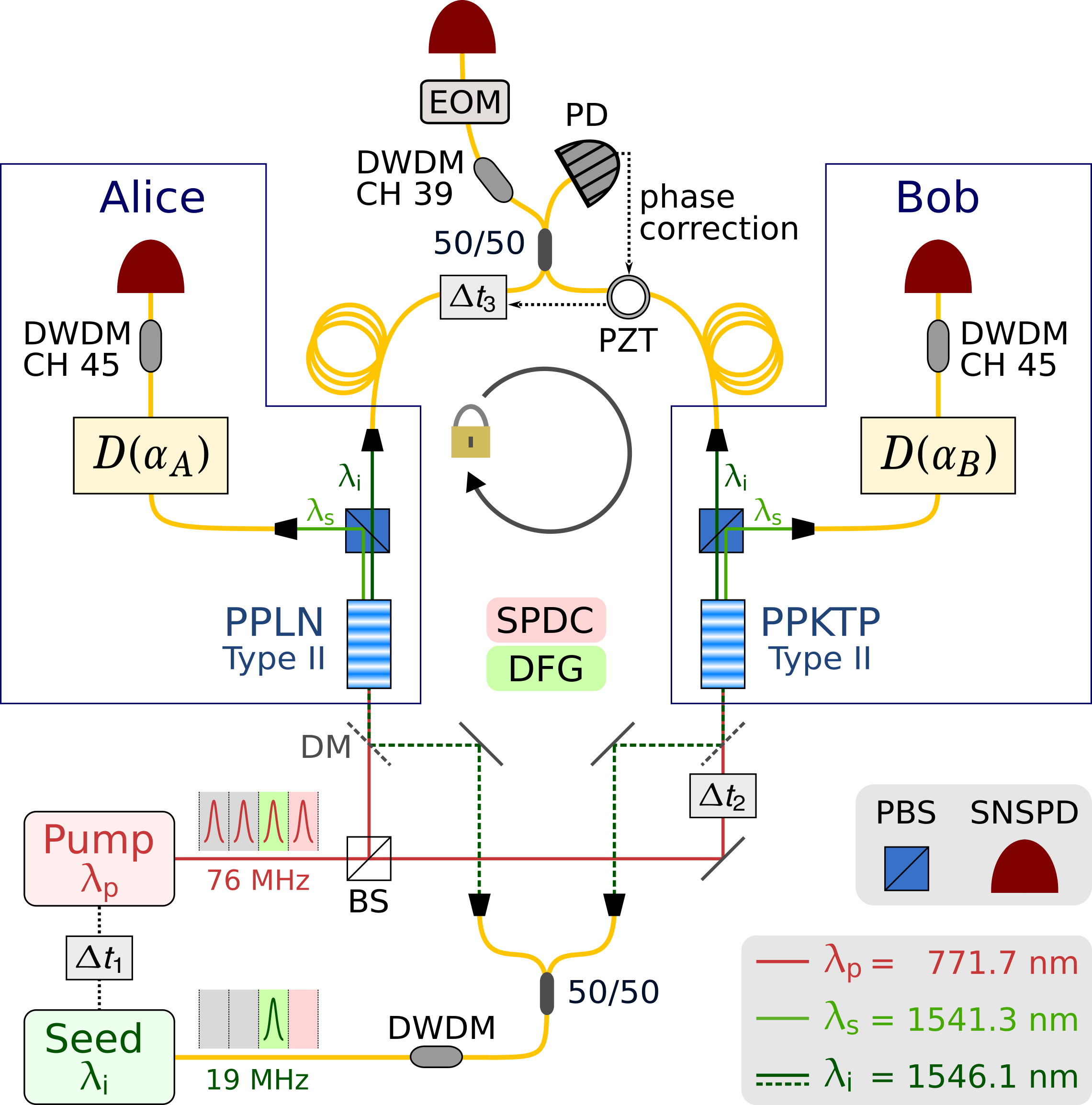}
\caption{\label{fig:Setup} Simplified schematic of the experimental setup for the heralded distribution and certification of single-photon path entanglement. A periodically poled lithium niobate (PPLN) and a periodically poled potassium titanyl phosphate (PPKTP) bulk nonlinear crystal are pumped by a pulsed Ti:sapphire laser at $\lambda_p$ in the picosecond regime with a repetition rate of \SI{76}{MHz} for collinear type-II SPDC and seeded by a pulsed DFB laser at $\lambda_i$ with a repetition rate of \SI{19}{MHz} for DFG. The idler photons are sent to the central station and herald entanglement distribution. The signal photons and coherent states are sent to Alice and Bob, respectively, in order to perform the displacement-based measurement. Photons are detected by three superconducting nanowire single-photon detectors (SNSPD).}
\end{figure}

 
\emph{Experiment.}---A schematic overview of the experimental implementation is presented in Fig.~\ref{fig:Setup}. We use two nonlinear crystals as type-II SPDC sources pumped by a pulsed laser at $\lambda_p=\SI{771.7}{nm}$ to create nondegenerate photons at $\lambda_s=\SI{1541.3}{nm}$ (signal) and $\lambda_i=\SI{1546.1}{nm}$ (idler). The photon pair creation probability per pump pulse for each crystal is kept at $P_\mathrm{pair} \approx \num{3E-3}$ in order to keep the probability of having double-pair emissions sufficiently low. Signal and idler modes are separated after their generation at the polarizing beam splitters (PBS) and coupled into single-mode optical fibers. 
The idler photons are then sent to a 50/50 beam splitter (BS) and are spectrally filtered by a dense wavelength division multiplexer (DWDM) with a \SI{100}{GHz} passband (ITU channel 39).  We therefore ensure high-purity heralded signal photons and achieve a spectral overlap of \SI{99.9}{\%} between idler photons originating from the two independent sources (see Supplemental Material Sec.~VIII).
To reduce the photon noise due to residual seed-pulse photons (see below), and unwanted optical reflections, arriving before the heralding idler photons, a gate of \SI{2}{ns} is generated by an electro-optic intensity modulator (EOM) to temporally filter before the detector. The EOM has an insertion loss of \SI{5.0}{dB} and an extinction ratio of \SI{33}{dB}. \par
In order to perform the displacement-based measurement on the state shared between Alice and Bob, we generate a coherent state with the same spectral, temporal and polarization properties as the single photon signal via a difference frequency generation (DFG) process by stimulating the nonlinear crystals with a pulsed distributed feedback (DFB) seed laser at a wavelength $\lambda_i=\SI{1546.1}{nm}$ and repetition rate \SI{19}{MHz}.
The seed laser is driven from well below to above the lasing threshold each cycle to phase randomize the coherent state. 
For the implementation of the displacement-based measurement (see Supplemental Material Sec.~VII), the single-photon and the coherent states are temporally brought to coincidence in an asymmetric Mach-Zehnder interferometer (AMZI) followed by a PBS to project the single-photon and the coherent states into the same temporal and polarization modes, which realizes the displacement operation~\cite{Paris1996}. We increase the spectral overlap between single-photon and coherent states by local filtering with DWDMs (\SI{100}{GHz} passband at ITU channel 45). \par

To fulfill the phase stability requirement (see Supplemental Material Sec.~VI), the central interferometer is phase locked using the residual seed laser pulses at the central station. A piezoelectric fiber stretcher (PZT) with a half-wavelength voltage of $V_\pi=\SI{0.18}{V}$ and an optical delay range of about \SI{0.57}{ps} is actively controlled such that the seed power at the second output port of the 50/50 BS, measured with a photodiode (PD), is maximized. Note that we do not require phase-coherent pump pulses (see Supplemental Material Sec.~VI). Additionally, and specific to our implementation of the displacement-based measurement, the phase difference between the AMZIs is stabilized (see Supplemental Material Sec.~VII). \par

\begin{figure}[tb]
\capstart
\includegraphics[height=4.5cm]{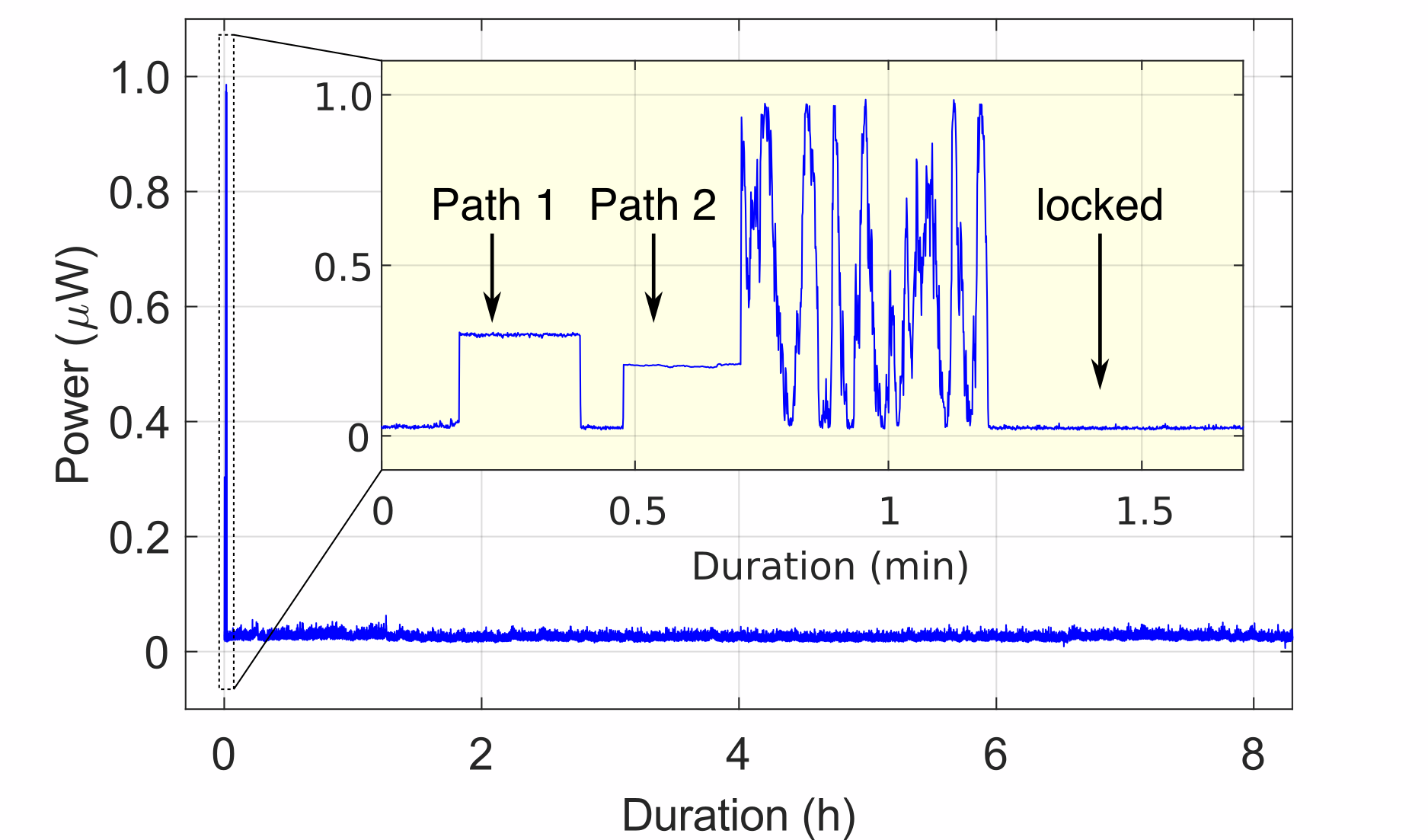}
\caption{\label{fig:PhaseStability} Characterization measurement of the central interferometer phase locking for $l=\SI{1.0}{km}$. The graph shows the measured seed power at the 50/50 BS output port 1 over a duration of \SI{8}{h}. The active feedback on the piezoelectric fiber stretcher and the time delay feedback are turned on after \SI{1.2}{min} (inset). During the initial \SI{1.2}{min}, first path 1 of the interferometer is left open only, then path 2, and afterwards both paths are opened leading to interference.}
\end{figure}

In order to demonstrate the feasibility of long-distance entanglement distribution, we extend the central interferometer arm lengths from initially $l=\SI{42}{m}$ to $l=\SI{1.0}{km}$ by inserting two fiber coils. This change additionally requires active polarization control before the 50/50 BS as well as active compensation of slow relative drifts in optical length between the two interferometer arms. 
Therefore, electronic polarization controllers (Phoenix Photonics PSC) are inserted after the fiber coils to minimize the seed power at the second output ports of the fiber PBSs whose first output ports are connected to the polarization maintaining 50/50 BS at the end of the central interferometer. 
The slow relative optical length drifts are compensated by actively setting $\Delta t_3$ (see Fig.~\ref{fig:Setup}) with a motorized delay line such that the voltage applied to the PZT is kept in range. 
In this way, we achieve long-term phase stabilization as shown in Fig.~\ref{fig:PhaseStability} for a duration of \SI{8}{h}. \par 

The photons are detected by three in-house-developed MoSi superconducting nanowire single-photon detectors (SNSPD) with efficiencies $\eta_d>\SI{60}{\%}$ and recovery times $\tau_\mathrm{rec}<\SI{35}{ns}$. Time correlated single-photon counting (ID Quantique ID900) is used to register the events of a signal photon detected by Alice, by Bob, and coincidences conditioned on the detection of an idler photon at the central station within a \SI{400}{ps} window with respect to the \SI{19}{MHz} clock signal. In the $\alpha$-basis, we monitor the displacement amplitudes by tracking the detection rates caused by coherent states arriving 1 cycle (\SI{52}{ns}) later than the expected signal photons. \par
\begin{figure}[tb]
\capstart
\includegraphics[height=4.5cm]{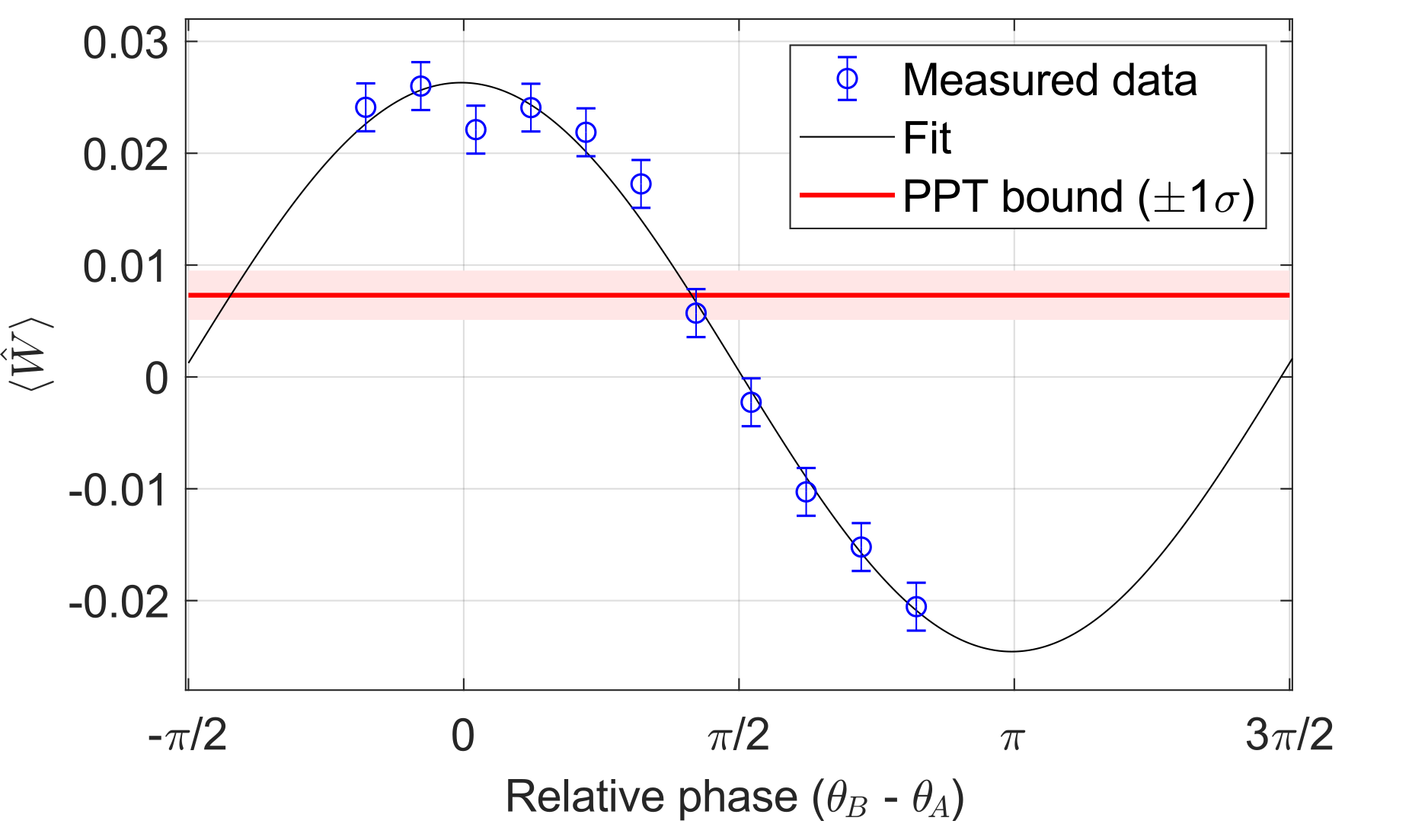}
\caption{\label{fig:Witness} Expectation value and PPT bound $w_\mathrm{ppt}^\mathrm{max}$ of the entanglement witness as a function of the relative phase $(\theta_B-\theta_A)$ between the displacement measurements for Alice and Bob with $l=\SI{1.0}{km}$. Values of $\langle\hat{W}\rangle$ larger than $w_\mathrm{ppt}^\mathrm{max}$ demonstrate entanglement. For each phase setting, counts were acquired for \SI{200}{s} in the $\alpha$-basis with $\alpha_1={0.818^{+0.004}_{-0.003}}$ and $\alpha_2={0.830^{+0.006}_{-0.007}}$ indicating the mean, maximum and minimum displacement amplitudes. Error bars of the measured data represent 1 standard deviation.}
\end{figure}

 
\emph{Results.}---A measurement of the witness as a function of the relative phase between Alice's and Bob's displacement operations is shown in Fig.~\ref{fig:Witness}. After the relative phase is set to $(\theta_B-\theta_A)=0$, counts were acquired in the $\alpha$-basis for \SI{1}{h} and subsequently in the $z$-basis for \SI{2.5}{h} by blocking the coherent state paths in the measurement interferometers. From the ``click''-``no-click'' events recorded by Alice and Bob, the corresponding joint probabilities are deduced (see Supplemental Material Sec.~IX). We separately determined the probability of having more than one photon locally in a Hanbury Brown-Twiss experiment for both Alice and Bob. Together with the joint probabilities measured in the $z$-basis as well as the displacement parameter amplitudes used in the $\alpha$-basis measurement, we compute the estimator (see Supplemental Material Sec.~V) for the maximal separable bound $w_\mathrm{ppt}^\mathrm{max}$ according to Eq.~\eqref{eq:wpptmax}. The experimental value for the expectation value $w_\rho^\mathrm{exp}$ of the witness $\hat{W}$ is computed from the measured joint probabilities in the $\alpha$-basis by
\begin{equation}
\label{eq:wrho}
w_\rho^\mathrm{exp} = (P_\mathrm{nc,nc}+P_\mathrm{c,c}-P_\mathrm{c,nc}-P_\mathrm{nc,c}) |_{\alpha_1,\alpha_2}.
\end{equation}
\par 
For the analysis of uncertainties on the separable bound $\sigma_\mathrm{ppt}^\mathrm{max}$ and on the experimental value of the witness $\sigma_\rho^\mathrm{exp}$ we assume the coincidence probabilities to be independent and identically distributed (i.i.d.) random variables (see Supplemental Material Sec.~V). The obtained results, as shown in Table~\ref{tab:ResultsWitness}, certify a violation of the entanglement witness by more than 5 standard deviations at a heralding rate of at least \SI{1.4}{kHz} for fiber distances of $l=\SI{42}{m}$ and $l=\SI{1.0}{km}$ inserted in each arm of the central interferometer. The higher total heralding rate in the case of $l=\SI{1.0}{km}$ is due to an elevated noise contribution. The larger statistical significance of the result with $l=\SI{1.0}{km}$ compared to $l=\SI{42}{m}$ is mainly attributed to a different alignment setting in the measurement AMZIs leading to an increase of the transmission on the entangled state (see Supplemental Material Sec.~IX). \par 
\begin{table}[b]
\capstart
\centering
\begin{ruledtabular}
\begin{tabular}{c c c c c c} 
$l$ 		& $\nu_h$ (kHz) 	& SNR 	& $w_\rho^\mathrm{exp}$ 	& $w_\mathrm{ppt}^\mathrm{max}$ 	& $k$ \\[0.5ex]
\cmidrule{1-6}
$\SI{42}{m}$ 	& 1.4 			& 12 		& 0.0206(8) 				& 0.0071(16) 				& 5.6 \\
$\SI{1.0}{km}$ 	& 1.6 			& 5 		& 0.0253(7) 				& 0.0071(22) 				& 6.2
\end{tabular}
\end{ruledtabular}
\caption{\label{tab:ResultsWitness} Measured value $w_\rho^\mathrm{exp}$ and calculated separable bound $w_\mathrm{ppt}^\mathrm{max}$ of the entanglement witness for fibers of length $l$ inserted in each arm of the central interferometer at an observed heralding rate $\nu_h$ and signal-to-noise ratio (SNR). The witness is violated by $k=(w_\rho^\mathrm{exp}-w_\mathrm{ppt}^\mathrm{max})/(\sigma_\mathrm{ppt}^\mathrm{max}+\sigma_\rho^\mathrm{exp})$ standard deviations.}
\end{table}
\emph{Discussion.}---Single-photon quantum repeater schemes are promising for fiber-based long-distance entanglement distribution because of their favorable transmission loss scaling, their robustness to memory and detector inefficiencies and the need for fewer resources than protocols based on two-photon detections~\cite{Sangouard2011}. We demonstrated the feasibility of such a scheme by actively stabilizing the phase of an interferometer with arm lengths of \SI{1.0}{km} and utilizing local displacement-based measurements for entanglement certification. In our scheme the phase difference between the AMZIs also needs to be stabilized (see Supplemental Material Sec.~VII), however, in a quantum repeater these AMZIs could be replaced, and the displacement performed, by quantum memories~\cite{Jobez2015,Kutluer2019}. \par
In principle, the scheme can be extended for real world applications by using two individual pump lasers, adding fiber before the sources to distribute the seed pulses and further increasing the size of the central interferometer. The main technical challenge in such an implementation is the increase of phase noise in a larger central interferometer~\cite{Minar2008}. This would not only degrade the entanglement, but also increase the leakage of residual seed laser pulse photons towards the heralding detector. Our solution to suppress them with an EOM introduces unwanted loss on the heralding photons, however, a better solution would be to develop a gated SNSPD for the central station. \par 
In conclusion, we demonstrated the heralded distribution of single-photon path entanglement in a repeater-like architecture. For a fiber distance of $2\!\times\!\SI{1.0}{km}$ inserted in the central interferometer we achieve a heralding rate of \SI{1.6}{kHz} and we certify a violation of the entanglement witness by \num{6.2} standard deviations. 
These results highlight the feasibility and challenges associated with realizing DLCZ-like quantum repeater architectures. \par 
\begin{acknowledgments}
The authors would like to thank C.~Autebert, J.-D.~Bancal, F.~Bussières, C.~Barreiro, A.~Boaron and P.~Remy for useful discussions and technical support. This work was supported by the Swiss National Science Foundation SNSF (Grants No.~200021\_159592, No.~200020\_182664, and No.~PP00P2-179109), COST (SBFI/SNF) IZCNZ0-174835, and the European Union's Horizon 2020 research and innovation programme under Grant Agreement No.~820445 (Quantum Internet Alliance).
\end{acknowledgments}

\bibliographystyle{apsrev4-1}
\bibliography{./references_arXiv_v2}

\begin{thebibliography}{26}%
\makeatletter
\providecommand \@ifxundefined [1]{%
 \@ifx{#1\undefined}
}%
\providecommand \@ifnum [1]{%
 \ifnum #1\expandafter \@firstoftwo
 \else \expandafter \@secondoftwo
 \fi
}%
\providecommand \@ifx [1]{%
 \ifx #1\expandafter \@firstoftwo
 \else \expandafter \@secondoftwo
 \fi
}%
\providecommand \natexlab [1]{#1}%
\providecommand \enquote  [1]{``#1''}%
\providecommand \bibnamefont  [1]{#1}%
\providecommand \bibfnamefont [1]{#1}%
\providecommand \citenamefont [1]{#1}%
\providecommand \href@noop [0]{\@secondoftwo}%
\providecommand \href [0]{\begingroup \@sanitize@url \@href}%
\providecommand \@href[1]{\@@startlink{#1}\@@href}%
\providecommand \@@href[1]{\endgroup#1\@@endlink}%
\providecommand \@sanitize@url [0]{\catcode `\\12\catcode `\$12\catcode
  `\&12\catcode `\#12\catcode `\^12\catcode `\_12\catcode `\%12\relax}%
\providecommand \@@startlink[1]{}%
\providecommand \@@endlink[0]{}%
\providecommand \url  [0]{\begingroup\@sanitize@url \@url }%
\providecommand \@url [1]{\endgroup\@href {#1}{\urlprefix }}%
\providecommand \urlprefix  [0]{URL }%
\providecommand \Eprint [0]{\href }%
\providecommand \doibase [0]{http://dx.doi.org/}%
\providecommand \selectlanguage [0]{\@gobble}%
\providecommand \bibinfo  [0]{\@secondoftwo}%
\providecommand \bibfield  [0]{\@secondoftwo}%
\providecommand \translation [1]{[#1]}%
\providecommand \BibitemOpen [0]{}%
\providecommand \bibitemStop [0]{}%
\providecommand \bibitemNoStop [0]{.\EOS\space}%
\providecommand \EOS [0]{\spacefactor3000\relax}%
\providecommand \BibitemShut  [1]{\csname bibitem#1\endcsname}%
\let\auto@bib@innerbib\@empty
\bibitem [{\citenamefont {Kimble}(2008)}]{Kimble2008}%
  \BibitemOpen
  \bibfield  {author} {\bibinfo {author} {\bibfnamefont {H.~J.}\ \bibnamefont
  {Kimble}},\ }\href {\doibase 10.1038/nature07127} {\bibfield  {journal}
  {\bibinfo  {journal} {Nature}\ }\textbf {\bibinfo {volume} {453}},\ \bibinfo
  {pages} {1023} (\bibinfo {year} {2008})}\BibitemShut {NoStop}%
\bibitem [{\citenamefont {Wehner}\ \emph {et~al.}(2018)\citenamefont {Wehner},
  \citenamefont {Elkouss},\ and\ \citenamefont {Hanson}}]{Wehner2018}%
  \BibitemOpen
  \bibfield  {author} {\bibinfo {author} {\bibfnamefont {S.}~\bibnamefont
  {Wehner}}, \bibinfo {author} {\bibfnamefont {D.}~\bibnamefont {Elkouss}}, \
  and\ \bibinfo {author} {\bibfnamefont {R.}~\bibnamefont {Hanson}},\
  }\href@noop {} {\bibfield  {journal} {\bibinfo  {journal} {Science}\ }\textbf
  {\bibinfo {volume} {362}},\ \bibinfo {pages} {eaam9288} (\bibinfo {year}
  {2018})}\BibitemShut {NoStop}%
\bibitem [{\citenamefont {Briegel}\ \emph {et~al.}(1998)\citenamefont
  {Briegel}, \citenamefont {D\"ur}, \citenamefont {Cirac},\ and\ \citenamefont
  {Zoller}}]{Briegel1998}%
  \BibitemOpen
  \bibfield  {author} {\bibinfo {author} {\bibfnamefont {H.-J.}\ \bibnamefont
  {Briegel}}, \bibinfo {author} {\bibfnamefont {W.}~\bibnamefont {D\"ur}},
  \bibinfo {author} {\bibfnamefont {J.~I.}\ \bibnamefont {Cirac}}, \ and\
  \bibinfo {author} {\bibfnamefont {P.}~\bibnamefont {Zoller}},\ }\href
  {\doibase 10.1103/PhysRevLett.81.5932} {\bibfield  {journal} {\bibinfo
  {journal} {Phys. Rev. Lett.}\ }\textbf {\bibinfo {volume} {81}},\ \bibinfo
  {pages} {5932} (\bibinfo {year} {1998})}\BibitemShut {NoStop}%
\bibitem [{\citenamefont {Tan}\ \emph {et~al.}(1991)\citenamefont {Tan},
  \citenamefont {Walls},\ and\ \citenamefont {Collett}}]{Tan1991}%
  \BibitemOpen
  \bibfield  {author} {\bibinfo {author} {\bibfnamefont {S.~M.}\ \bibnamefont
  {Tan}}, \bibinfo {author} {\bibfnamefont {D.~F.}\ \bibnamefont {Walls}}, \
  and\ \bibinfo {author} {\bibfnamefont {M.~J.}\ \bibnamefont {Collett}},\
  }\href {\doibase 10.1103/PhysRevLett.66.252} {\bibfield  {journal} {\bibinfo
  {journal} {Phys. Rev. Lett.}\ }\textbf {\bibinfo {volume} {66}},\ \bibinfo
  {pages} {252} (\bibinfo {year} {1991})}\BibitemShut {NoStop}%
\bibitem [{\citenamefont {Sangouard}\ \emph {et~al.}(2011)\citenamefont
  {Sangouard}, \citenamefont {Simon}, \citenamefont {de~Riedmatten},\ and\
  \citenamefont {Gisin}}]{Sangouard2011}%
  \BibitemOpen
  \bibfield  {author} {\bibinfo {author} {\bibfnamefont {N.}~\bibnamefont
  {Sangouard}}, \bibinfo {author} {\bibfnamefont {C.}~\bibnamefont {Simon}},
  \bibinfo {author} {\bibfnamefont {H.}~\bibnamefont {de~Riedmatten}}, \ and\
  \bibinfo {author} {\bibfnamefont {N.}~\bibnamefont {Gisin}},\ }\href
  {\doibase 10.1103/RevModPhys.83.33} {\bibfield  {journal} {\bibinfo
  {journal} {Rev. Mod. Phys.}\ }\textbf {\bibinfo {volume} {83}},\ \bibinfo
  {pages} {33} (\bibinfo {year} {2011})}\BibitemShut {NoStop}%
\bibitem [{\citenamefont {Duan}\ \emph {et~al.}(2001)\citenamefont {Duan},
  \citenamefont {Lukin}, \citenamefont {Cirac},\ and\ \citenamefont
  {Zoller}}]{Duan2001}%
  \BibitemOpen
  \bibfield  {author} {\bibinfo {author} {\bibfnamefont {L.-M.}\ \bibnamefont
  {Duan}}, \bibinfo {author} {\bibfnamefont {M.~D.}\ \bibnamefont {Lukin}},
  \bibinfo {author} {\bibfnamefont {J.~I.}\ \bibnamefont {Cirac}}, \ and\
  \bibinfo {author} {\bibfnamefont {P.}~\bibnamefont {Zoller}},\ }\href
  {\doibase 10.1038/35106500} {\bibfield  {journal} {\bibinfo  {journal}
  {Nature}\ }\textbf {\bibinfo {volume} {414}},\ \bibinfo {pages} {413}
  (\bibinfo {year} {2001})}\BibitemShut {NoStop}%
\bibitem [{\citenamefont {Simon}\ \emph {et~al.}(2007)\citenamefont {Simon},
  \citenamefont {de~Riedmatten}, \citenamefont {Afzelius}, \citenamefont
  {Sangouard}, \citenamefont {Zbinden},\ and\ \citenamefont
  {Gisin}}]{Simon2007}%
  \BibitemOpen
  \bibfield  {author} {\bibinfo {author} {\bibfnamefont {C.}~\bibnamefont
  {Simon}}, \bibinfo {author} {\bibfnamefont {H.}~\bibnamefont
  {de~Riedmatten}}, \bibinfo {author} {\bibfnamefont {M.}~\bibnamefont
  {Afzelius}}, \bibinfo {author} {\bibfnamefont {N.}~\bibnamefont {Sangouard}},
  \bibinfo {author} {\bibfnamefont {H.}~\bibnamefont {Zbinden}}, \ and\
  \bibinfo {author} {\bibfnamefont {N.}~\bibnamefont {Gisin}},\ }\href
  {\doibase 10.1103/PhysRevLett.98.190503} {\bibfield  {journal} {\bibinfo
  {journal} {Phys. Rev. Lett.}\ }\textbf {\bibinfo {volume} {98}},\ \bibinfo
  {pages} {190503} (\bibinfo {year} {2007})}\BibitemShut {NoStop}%
\bibitem [{\citenamefont {Slodi\ifmmode~\check{c}\else \v{c}\fi{}ka}\ \emph
  {et~al.}(2013)\citenamefont {Slodi\ifmmode~\check{c}\else \v{c}\fi{}ka},
  \citenamefont {H\'etet}, \citenamefont {R\"ock}, \citenamefont {Schindler},
  \citenamefont {Hennrich},\ and\ \citenamefont {Blatt}}]{Slodicka2013}%
  \BibitemOpen
  \bibfield  {author} {\bibinfo {author} {\bibfnamefont {L.}~\bibnamefont
  {Slodi\ifmmode~\check{c}\else \v{c}\fi{}ka}}, \bibinfo {author}
  {\bibfnamefont {G.}~\bibnamefont {H\'etet}}, \bibinfo {author} {\bibfnamefont
  {N.}~\bibnamefont {R\"ock}}, \bibinfo {author} {\bibfnamefont
  {P.}~\bibnamefont {Schindler}}, \bibinfo {author} {\bibfnamefont
  {M.}~\bibnamefont {Hennrich}}, \ and\ \bibinfo {author} {\bibfnamefont
  {R.}~\bibnamefont {Blatt}},\ }\href {\doibase 10.1103/PhysRevLett.110.083603}
  {\bibfield  {journal} {\bibinfo  {journal} {Phys. Rev. Lett.}\ }\textbf
  {\bibinfo {volume} {110}},\ \bibinfo {pages} {083603} (\bibinfo {year}
  {2013})}\BibitemShut {NoStop}%
\bibitem [{\citenamefont {Delteil}\ \emph {et~al.}(2016)\citenamefont
  {Delteil}, \citenamefont {Sun}, \citenamefont {Gao}, \citenamefont {Togan},
  \citenamefont {Faelt},\ and\ \citenamefont {Imamoglu}}]{Delteil2016}%
  \BibitemOpen
  \bibfield  {author} {\bibinfo {author} {\bibfnamefont {A.}~\bibnamefont
  {Delteil}}, \bibinfo {author} {\bibfnamefont {Z.}~\bibnamefont {Sun}},
  \bibinfo {author} {\bibfnamefont {W.~B.}\ \bibnamefont {Gao}}, \bibinfo
  {author} {\bibfnamefont {E.}~\bibnamefont {Togan}}, \bibinfo {author}
  {\bibfnamefont {S.}~\bibnamefont {Faelt}}, \ and\ \bibinfo {author}
  {\bibfnamefont {A.}~\bibnamefont {Imamoglu}},\ }\href {\doibase
  10.1038/nphys3605} {\bibfield  {journal} {\bibinfo  {journal} {Nature
  Physics}\ }\textbf {\bibinfo {volume} {12}},\ \bibinfo {pages} {218}
  (\bibinfo {year} {2016})}\BibitemShut {NoStop}%
\bibitem [{\citenamefont {Stockill}\ \emph {et~al.}(2017)\citenamefont
  {Stockill}, \citenamefont {Stanley}, \citenamefont {Huthmacher},
  \citenamefont {Clarke}, \citenamefont {Hugues}, \citenamefont {Miller},
  \citenamefont {Matthiesen}, \citenamefont {Le~Gall},\ and\ \citenamefont
  {Atat\"ure}}]{Stockill2017}%
  \BibitemOpen
  \bibfield  {author} {\bibinfo {author} {\bibfnamefont {R.}~\bibnamefont
  {Stockill}}, \bibinfo {author} {\bibfnamefont {M.~J.}\ \bibnamefont
  {Stanley}}, \bibinfo {author} {\bibfnamefont {L.}~\bibnamefont {Huthmacher}},
  \bibinfo {author} {\bibfnamefont {E.}~\bibnamefont {Clarke}}, \bibinfo
  {author} {\bibfnamefont {M.}~\bibnamefont {Hugues}}, \bibinfo {author}
  {\bibfnamefont {A.~J.}\ \bibnamefont {Miller}}, \bibinfo {author}
  {\bibfnamefont {C.}~\bibnamefont {Matthiesen}}, \bibinfo {author}
  {\bibfnamefont {C.}~\bibnamefont {Le~Gall}}, \ and\ \bibinfo {author}
  {\bibfnamefont {M.}~\bibnamefont {Atat\"ure}},\ }\href {\doibase
  10.1103/PhysRevLett.119.010503} {\bibfield  {journal} {\bibinfo  {journal}
  {Phys. Rev. Lett.}\ }\textbf {\bibinfo {volume} {119}},\ \bibinfo {pages}
  {010503} (\bibinfo {year} {2017})}\BibitemShut {NoStop}%
\bibitem [{\citenamefont {Humphreys}\ \emph {et~al.}(2018)\citenamefont
  {Humphreys}, \citenamefont {Kalb}, \citenamefont {Morits}, \citenamefont
  {Schouten}, \citenamefont {Vermeulen}, \citenamefont {Twitchen},
  \citenamefont {Markham},\ and\ \citenamefont {Hanson}}]{Humphreys2018}%
  \BibitemOpen
  \bibfield  {author} {\bibinfo {author} {\bibfnamefont {P.~C.}\ \bibnamefont
  {Humphreys}}, \bibinfo {author} {\bibfnamefont {N.}~\bibnamefont {Kalb}},
  \bibinfo {author} {\bibfnamefont {J.~P.~J.}\ \bibnamefont {Morits}}, \bibinfo
  {author} {\bibfnamefont {R.~N.}\ \bibnamefont {Schouten}}, \bibinfo {author}
  {\bibfnamefont {R.~F.~L.}\ \bibnamefont {Vermeulen}}, \bibinfo {author}
  {\bibfnamefont {D.~J.}\ \bibnamefont {Twitchen}}, \bibinfo {author}
  {\bibfnamefont {M.}~\bibnamefont {Markham}}, \ and\ \bibinfo {author}
  {\bibfnamefont {R.}~\bibnamefont {Hanson}},\ }\href {\doibase
  10.1038/s41586-018-0200-5} {\bibfield  {journal} {\bibinfo  {journal}
  {Nature}\ }\textbf {\bibinfo {volume} {558}},\ \bibinfo {pages} {268}
  (\bibinfo {year} {2018})}\BibitemShut {NoStop}%
\bibitem [{\citenamefont {Yu}\ \emph {et~al.}(2020)\citenamefont {Yu},
  \citenamefont {Ma}, \citenamefont {Luo}, \citenamefont {Jing}, \citenamefont
  {Sun}, \citenamefont {Fang}, \citenamefont {Yang}, \citenamefont {Liu},
  \citenamefont {Zheng}, \citenamefont {Xie}, \citenamefont {Zhang},
  \citenamefont {You}, \citenamefont {Wang}, \citenamefont {Chen},
  \citenamefont {Zhang}, \citenamefont {Bao},\ and\ \citenamefont
  {Pan}}]{Yu2020}%
  \BibitemOpen
  \bibfield  {author} {\bibinfo {author} {\bibfnamefont {Y.}~\bibnamefont
  {Yu}}, \bibinfo {author} {\bibfnamefont {F.}~\bibnamefont {Ma}}, \bibinfo
  {author} {\bibfnamefont {X.-Y.}\ \bibnamefont {Luo}}, \bibinfo {author}
  {\bibfnamefont {B.}~\bibnamefont {Jing}}, \bibinfo {author} {\bibfnamefont
  {P.-F.}\ \bibnamefont {Sun}}, \bibinfo {author} {\bibfnamefont {R.-Z.}\
  \bibnamefont {Fang}}, \bibinfo {author} {\bibfnamefont {C.-W.}\ \bibnamefont
  {Yang}}, \bibinfo {author} {\bibfnamefont {H.}~\bibnamefont {Liu}}, \bibinfo
  {author} {\bibfnamefont {M.-Y.}\ \bibnamefont {Zheng}}, \bibinfo {author}
  {\bibfnamefont {X.-P.}\ \bibnamefont {Xie}}, \bibinfo {author} {\bibfnamefont
  {W.-J.}\ \bibnamefont {Zhang}}, \bibinfo {author} {\bibfnamefont {L.-X.}\
  \bibnamefont {You}}, \bibinfo {author} {\bibfnamefont {Z.}~\bibnamefont
  {Wang}}, \bibinfo {author} {\bibfnamefont {T.-y.}\ \bibnamefont {Chen}},
  \bibinfo {author} {\bibfnamefont {Q.}~\bibnamefont {Zhang}}, \bibinfo
  {author} {\bibfnamefont {X.-H.}\ \bibnamefont {Bao}}, \ and\ \bibinfo
  {author} {\bibfnamefont {J.-W.}\ \bibnamefont {Pan}},\ }\href {\doibase
  10.1038/s41586-020-1976-7} {\bibfield  {journal} {\bibinfo  {journal}
  {Nature}\ }\textbf {\bibinfo {volume} {578}},\ \bibinfo {pages} {240}
  (\bibinfo {year} {2020})}\BibitemShut {NoStop}%
\bibitem [{\citenamefont {Monteiro}\ \emph {et~al.}(2015)\citenamefont
  {Monteiro}, \citenamefont {Vivoli}, \citenamefont {Guerreiro}, \citenamefont
  {Martin}, \citenamefont {Bancal}, \citenamefont {Zbinden}, \citenamefont
  {Thew},\ and\ \citenamefont {Sangouard}}]{Monteiro2015}%
  \BibitemOpen
  \bibfield  {author} {\bibinfo {author} {\bibfnamefont {F.}~\bibnamefont
  {Monteiro}}, \bibinfo {author} {\bibfnamefont {V.~C.}\ \bibnamefont
  {Vivoli}}, \bibinfo {author} {\bibfnamefont {T.}~\bibnamefont {Guerreiro}},
  \bibinfo {author} {\bibfnamefont {A.}~\bibnamefont {Martin}}, \bibinfo
  {author} {\bibfnamefont {J.-D.}\ \bibnamefont {Bancal}}, \bibinfo {author}
  {\bibfnamefont {H.}~\bibnamefont {Zbinden}}, \bibinfo {author} {\bibfnamefont
  {R.~T.}\ \bibnamefont {Thew}}, \ and\ \bibinfo {author} {\bibfnamefont
  {N.}~\bibnamefont {Sangouard}},\ }\href {\doibase
  10.1103/PhysRevLett.114.170504} {\bibfield  {journal} {\bibinfo  {journal}
  {Phys. Rev. Lett.}\ }\textbf {\bibinfo {volume} {114}},\ \bibinfo {pages}
  {170504} (\bibinfo {year} {2015})}\BibitemShut {NoStop}%
\bibitem [{\citenamefont {Vivoli}\ \emph {et~al.}(2015)\citenamefont {Vivoli},
  \citenamefont {Sekatski}, \citenamefont {Bancal}, \citenamefont {Lim},
  \citenamefont {Martin}, \citenamefont {Thew}, \citenamefont {Zbinden},
  \citenamefont {Gisin},\ and\ \citenamefont {Sangouard}}]{Vivoli2015}%
  \BibitemOpen
  \bibfield  {author} {\bibinfo {author} {\bibfnamefont {V.~C.}\ \bibnamefont
  {Vivoli}}, \bibinfo {author} {\bibfnamefont {P.}~\bibnamefont {Sekatski}},
  \bibinfo {author} {\bibfnamefont {J.-D.}\ \bibnamefont {Bancal}}, \bibinfo
  {author} {\bibfnamefont {C.~C.~W.}\ \bibnamefont {Lim}}, \bibinfo {author}
  {\bibfnamefont {A.}~\bibnamefont {Martin}}, \bibinfo {author} {\bibfnamefont
  {R.~T.}\ \bibnamefont {Thew}}, \bibinfo {author} {\bibfnamefont
  {H.}~\bibnamefont {Zbinden}}, \bibinfo {author} {\bibfnamefont
  {N.}~\bibnamefont {Gisin}}, \ and\ \bibinfo {author} {\bibfnamefont
  {N.}~\bibnamefont {Sangouard}},\ }\href {\doibase
  10.1088/1367-2630/17/2/023023} {\bibfield  {journal} {\bibinfo  {journal}
  {New Journal of Physics}\ }\textbf {\bibinfo {volume} {17}},\ \bibinfo
  {pages} {023023} (\bibinfo {year} {2015})}\BibitemShut {NoStop}%
\bibitem [{\citenamefont {Paris}(1996)}]{Paris1996}%
  \BibitemOpen
  \bibfield  {author} {\bibinfo {author} {\bibfnamefont {M.~G.}\ \bibnamefont
  {Paris}},\ }\href {\doibase 10.1016/0375-9601(96)00339-8} {\bibfield
  {journal} {\bibinfo  {journal} {Physics Letters A}\ }\textbf {\bibinfo
  {volume} {217}},\ \bibinfo {pages} {78} (\bibinfo {year} {1996})}\BibitemShut
  {NoStop}%
\bibitem [{\citenamefont {Banaszek}\ and\ \citenamefont
  {W\'odkiewicz}(1999)}]{Banaszek1999}%
  \BibitemOpen
  \bibfield  {author} {\bibinfo {author} {\bibfnamefont {K.}~\bibnamefont
  {Banaszek}}\ and\ \bibinfo {author} {\bibfnamefont {K.}~\bibnamefont
  {W\'odkiewicz}},\ }\href {\doibase 10.1103/PhysRevLett.82.2009} {\bibfield
  {journal} {\bibinfo  {journal} {Phys. Rev. Lett.}\ }\textbf {\bibinfo
  {volume} {82}},\ \bibinfo {pages} {2009} (\bibinfo {year}
  {1999})}\BibitemShut {NoStop}%
\bibitem [{\citenamefont {Kuzmich}\ \emph {et~al.}(2000)\citenamefont
  {Kuzmich}, \citenamefont {Walmsley},\ and\ \citenamefont
  {Mandel}}]{Kuzmich2000}%
  \BibitemOpen
  \bibfield  {author} {\bibinfo {author} {\bibfnamefont {A.}~\bibnamefont
  {Kuzmich}}, \bibinfo {author} {\bibfnamefont {I.~A.}\ \bibnamefont
  {Walmsley}}, \ and\ \bibinfo {author} {\bibfnamefont {L.}~\bibnamefont
  {Mandel}},\ }\href {\doibase 10.1103/PhysRevLett.85.1349} {\bibfield
  {journal} {\bibinfo  {journal} {Phys. Rev. Lett.}\ }\textbf {\bibinfo
  {volume} {85}},\ \bibinfo {pages} {1349} (\bibinfo {year}
  {2000})}\BibitemShut {NoStop}%
\bibitem [{\citenamefont {Bj{\"o}rk}\ \emph {et~al.}(2001)\citenamefont
  {Bj{\"o}rk}, \citenamefont {Jonsson},\ and\ \citenamefont
  {S{\'a}nchez-Soto}}]{Bjork2001}%
  \BibitemOpen
  \bibfield  {author} {\bibinfo {author} {\bibfnamefont {G.}~\bibnamefont
  {Bj{\"o}rk}}, \bibinfo {author} {\bibfnamefont {P.}~\bibnamefont {Jonsson}},
  \ and\ \bibinfo {author} {\bibfnamefont {L.~L.}\ \bibnamefont
  {S{\'a}nchez-Soto}},\ }\href {\doibase 10.1103/PhysRevA.64.042106} {\bibfield
   {journal} {\bibinfo  {journal} {Phys. Rev. A}\ }\textbf {\bibinfo {volume}
  {64}},\ \bibinfo {pages} {042106} (\bibinfo {year} {2001})}\BibitemShut
  {NoStop}%
\bibitem [{\citenamefont {Hessmo}\ \emph {et~al.}(2004)\citenamefont {Hessmo},
  \citenamefont {Usachev}, \citenamefont {Heydari},\ and\ \citenamefont
  {Bj\"ork}}]{Hessmo2004}%
  \BibitemOpen
  \bibfield  {author} {\bibinfo {author} {\bibfnamefont {B.}~\bibnamefont
  {Hessmo}}, \bibinfo {author} {\bibfnamefont {P.}~\bibnamefont {Usachev}},
  \bibinfo {author} {\bibfnamefont {H.}~\bibnamefont {Heydari}}, \ and\
  \bibinfo {author} {\bibfnamefont {G.}~\bibnamefont {Bj\"ork}},\ }\href
  {\doibase 10.1103/PhysRevLett.92.180401} {\bibfield  {journal} {\bibinfo
  {journal} {Phys. Rev. Lett.}\ }\textbf {\bibinfo {volume} {92}},\ \bibinfo
  {pages} {180401} (\bibinfo {year} {2004})}\BibitemShut {NoStop}%
\bibitem [{\citenamefont {Horodecki}\ \emph {et~al.}(1996)\citenamefont
  {Horodecki}, \citenamefont {Horodecki},\ and\ \citenamefont
  {Horodecki}}]{Horodecki1996}%
  \BibitemOpen
  \bibfield  {author} {\bibinfo {author} {\bibfnamefont {M.}~\bibnamefont
  {Horodecki}}, \bibinfo {author} {\bibfnamefont {P.}~\bibnamefont
  {Horodecki}}, \ and\ \bibinfo {author} {\bibfnamefont {R.}~\bibnamefont
  {Horodecki}},\ }\href {\doibase 10.1016/S0375-9601(96)00706-2} {\bibfield
  {journal} {\bibinfo  {journal} {Physics Letters A}\ }\textbf {\bibinfo
  {volume} {223}},\ \bibinfo {pages} {1} (\bibinfo {year} {1996})}\BibitemShut
  {NoStop}%
\bibitem [{\citenamefont {Jobez}(2015)}]{Jobez2015}%
  \BibitemOpen
  \bibfield  {author} {\bibinfo {author} {\bibfnamefont {P.}~\bibnamefont
  {Jobez}},\ }\emph {\bibinfo {title} {Stockage multimode au niveau quantique
  pendant une milliseconde}},\ \href
  {https://nbn-resolving.org/urn:nbn:ch:unige-836717} {Ph.D. thesis},\ \bibinfo
   {school} {Université de Genève} (\bibinfo {year} {2015})\BibitemShut
  {NoStop}%
\bibitem [{\citenamefont {Kutluer}\ \emph {et~al.}(2019)\citenamefont
  {Kutluer}, \citenamefont {Distante}, \citenamefont {Casabone}, \citenamefont
  {Duranti}, \citenamefont {Mazzera},\ and\ \citenamefont
  {de~Riedmatten}}]{Kutluer2019}%
  \BibitemOpen
  \bibfield  {author} {\bibinfo {author} {\bibfnamefont {K.}~\bibnamefont
  {Kutluer}}, \bibinfo {author} {\bibfnamefont {E.}~\bibnamefont {Distante}},
  \bibinfo {author} {\bibfnamefont {B.}~\bibnamefont {Casabone}}, \bibinfo
  {author} {\bibfnamefont {S.}~\bibnamefont {Duranti}}, \bibinfo {author}
  {\bibfnamefont {M.}~\bibnamefont {Mazzera}}, \ and\ \bibinfo {author}
  {\bibfnamefont {H.}~\bibnamefont {de~Riedmatten}},\ }\href {\doibase
  10.1103/PhysRevLett.123.030501} {\bibfield  {journal} {\bibinfo  {journal}
  {Phys. Rev. Lett.}\ }\textbf {\bibinfo {volume} {123}},\ \bibinfo {pages}
  {030501} (\bibinfo {year} {2019})}\BibitemShut {NoStop}%
\bibitem [{\citenamefont {Min{\'a}{\v{r}}}\ \emph {et~al.}(2008)\citenamefont
  {Min{\'a}{\v{r}}}, \citenamefont {de~Riedmatten}, \citenamefont {Simon},
  \citenamefont {Zbinden},\ and\ \citenamefont {Gisin}}]{Minar2008}%
  \BibitemOpen
  \bibfield  {author} {\bibinfo {author} {\bibfnamefont {J.}~\bibnamefont
  {Min{\'a}{\v{r}}}}, \bibinfo {author} {\bibfnamefont {H.}~\bibnamefont
  {de~Riedmatten}}, \bibinfo {author} {\bibfnamefont {C.}~\bibnamefont
  {Simon}}, \bibinfo {author} {\bibfnamefont {H.}~\bibnamefont {Zbinden}}, \
  and\ \bibinfo {author} {\bibfnamefont {N.}~\bibnamefont {Gisin}},\ }\href
  {\doibase 10.1103/PhysRevA.77.052325} {\bibfield  {journal} {\bibinfo
  {journal} {Phys. Rev. A}\ }\textbf {\bibinfo {volume} {77}},\ \bibinfo
  {pages} {052325} (\bibinfo {year} {2008})}\BibitemShut {NoStop}%
\bibitem [{\citenamefont {McMillan}\ \emph {et~al.}(2013)\citenamefont
  {McMillan}, \citenamefont {Labont{\'{e}}}, \citenamefont {Clark},
  \citenamefont {Bell}, \citenamefont {Alibart}, \citenamefont {Martin},
  \citenamefont {Wadsworth}, \citenamefont {Tanzilli},\ and\ \citenamefont
  {Rarity}}]{McMillan2013}%
  \BibitemOpen
  \bibfield  {author} {\bibinfo {author} {\bibfnamefont {A.~R.}\ \bibnamefont
  {McMillan}}, \bibinfo {author} {\bibfnamefont {L.}~\bibnamefont
  {Labont{\'{e}}}}, \bibinfo {author} {\bibfnamefont {A.~S.}\ \bibnamefont
  {Clark}}, \bibinfo {author} {\bibfnamefont {B.}~\bibnamefont {Bell}},
  \bibinfo {author} {\bibfnamefont {O.}~\bibnamefont {Alibart}}, \bibinfo
  {author} {\bibfnamefont {A.}~\bibnamefont {Martin}}, \bibinfo {author}
  {\bibfnamefont {W.~J.}\ \bibnamefont {Wadsworth}}, \bibinfo {author}
  {\bibfnamefont {S.}~\bibnamefont {Tanzilli}}, \ and\ \bibinfo {author}
  {\bibfnamefont {J.~G.}\ \bibnamefont {Rarity}},\ }\href {\doibase
  10.1038/srep02032} {\bibfield  {journal} {\bibinfo  {journal} {Scientific
  Reports}\ }\textbf {\bibinfo {volume} {3}},\ \bibinfo {pages} {2032}
  (\bibinfo {year} {2013})}\BibitemShut {NoStop}%
\bibitem [{\citenamefont {Mosley}(2007)}]{Mosley2007}%
  \BibitemOpen
  \bibfield  {author} {\bibinfo {author} {\bibfnamefont {P.~J.}\ \bibnamefont
  {Mosley}},\ }\emph {\bibinfo {title} {Generation of Heralded Single Photons
  in Pure Quantum States}},\ \href
  {https://ora.ox.ac.uk/objects/uuid:44c36e1e-11ee-41e2-ba29-611c932ce4ff}
  {Ph.D. thesis},\ \bibinfo  {school} {University of Oxford} (\bibinfo {year}
  {2007})\BibitemShut {NoStop}%
\bibitem [{\citenamefont {Bruno}\ \emph {et~al.}(2014)\citenamefont {Bruno},
  \citenamefont {Martin},\ and\ \citenamefont {Thew}}]{Bruno2014}%
  \BibitemOpen
  \bibfield  {author} {\bibinfo {author} {\bibfnamefont {N.}~\bibnamefont
  {Bruno}}, \bibinfo {author} {\bibfnamefont {A.}~\bibnamefont {Martin}}, \
  and\ \bibinfo {author} {\bibfnamefont {R.}~\bibnamefont {Thew}},\ }\href
  {\doibase 10.1016/j.optcom.2014.02.025} {\bibfield  {journal} {\bibinfo
  {journal} {Optics Communications}\ }\textbf {\bibinfo {volume} {327}},\
  \bibinfo {pages} {17} (\bibinfo {year} {2014})}\BibitemShut {NoStop}%
\end{thebibliography}%

\pagebreak
\clearpage
\onecolumngrid
{\bf\large\centering Supplemental Material: Heralded distribution of single-photon path entanglement \par}
\section{Entanglement witness}
\label{sec:app_witness}
We consider a scenario where we have two protagonists, Alice and Bob, each of whom has a photonic mode labelled 1 and 2, respectively. We introduce the corresponding bosonic operators $a_i$ and $a_i^{\dagger}$ with $i \in \{1,2\}$. We consider the observable 
\begin{equation}
\label{observable}
\mathcal{\hat{W}}=\sigma_{\alpha_1}^{(1)}\otimes\sigma_{\alpha_2}^{(2)}
\end{equation}
where $\sigma_{\alpha_i}^{(i)}$ is realized on mode $i$ with a single photon detector (non-photon-number-resolving detector) preceded by a displacement operation $D(\alpha)=e^{\alpha a_i^{\dagger}-\alpha^\ast a_i}$. The explicit expression of $\sigma_{\alpha_i}^{(i)}$ is given by
\begin{equation}
\sigma_{\alpha_i}^{(i)}=D^{\dagger}(\alpha_i)
(2\ket{0}\bra{0}-\mathds{1})D(\alpha_i)
\end{equation}
if one assigns the outcomes $+1$ when the detector does not click and $-1$ if it clicks. Moreover we consider the phase averaging of our observable according to 
\begin{equation}
\label{average}
\hat{W}=\frac{1}{2\pi}\int_0^{2\pi} d\phi\left(\prod_{i}^2 e^{i \phi a^{\dagger}_i a_i}\right)\mathcal{\hat{W}} \left(\prod_{i}^2 e^{-i \phi a^{\dagger}_i a_i}\right).
\end{equation}
Additionally we perform measurements without displacement and thus have access to the probabilities $P_\mathrm{nc,nc},$ $P_\mathrm{c,nc},$ $P_\mathrm{nc,c},$ $P_\mathrm{c,c},$ where for example $P_\mathrm{nc,c}$ is the probability that the first detector on mode 1 does not click and the second one mode 2 clicks.

\section{Separable bound in qubit space}
\label{sec:app_bound1}
We consider the qubit space $\{\ket{0},\ket{1}\}$ made with the vaccum and the single photon number state.
We want to find the maximum value of $\tr(\rho_{\mathrm{qubit}}^{\mathrm{sep}} \hat{W}_{\mathrm{qubit}})$ a separable 2-qubit state $\rho_\mathrm{qubit}^{\mathrm{sep}}$ can achieve. We use the Peres-Horodecki criterion (also refered to as PPT criterion) stating that any two qubit state having a positive partial transpose (PPT state) is separable and hence compute the maximum value of $\tr(\rho_{\mathrm{qubit}}^{\mathrm{sep}} \hat{W}_{\mathrm{qubit}})$ a PPT qubit state can achieve knowing the diagonal elements of $\rho_{\mathrm{qubit}}^{\mathrm{sep}}$. Let us note that $\hat{W}$ have a simple structure in this qubit space
\begin{equation}
\hat{W}_\mathrm{qubit}=
\quad
\begin{pmatrix} 
w_{00} 	& 0 		& 0 		& 0 \\
0 		& w_{01} 	& wc_{01} 	& 0  \\
0 		& wc_{10} 	& w_{10} 	& 0  \\
0 		& 0 		& 0 		& w_{11}  \\
\end{pmatrix}.
\end{equation} 
We can thus write 
\begin{equation}
\tr(\rho_\mathrm{qubit}^{\mathrm{sep}} \hat{W}_\mathrm{qubit}) = w_{00}P_{00}+w_{01}P_{01}+ w_{10}P_{10}+w_{11}P_{11}+wc_{01}\bra{10}\rho_\mathrm{qubit}^{\mathrm{sep}}\ket{01} + wc_{10}\bra{01}\rho_\mathrm{qubit}^{\mathrm{sep}}\ket{10}
\end{equation}
where $P_{ij} = \bra{ij}\rho_\mathrm{qubit}^{\mathrm{sep}}\ket{ij}$ represents the probability to get $i$ photons in mode 1 and $j$ photons in mode 2. \newline

Furthermore, the positivity of $\rho_\mathrm{qubit}^{\mathrm{sep}}$ implies that $|\bra{01}\rho_\mathrm{qubit}^{\mathrm{sep}}\ket{10}|\leq\sqrt{P_{10} P_{01}}$. The positivity under partial transpose imposes that $|\bra{01}\rho_\mathrm{qubit}^{\mathrm{sep}}\ket{10}|\leq\sqrt{P_{00} P_{11}}$. The condition ``$\rho_\mathrm{qubit}^{\mathrm{sep}}$ is a PPT state'' thus imposes that 
\begin{equation}
\tr(\rho_\mathrm{qubit}^{\mathrm{sep}} \hat{W}_\mathrm{qubit}) \leq w_{00}P_{00}+w_{01}P_{01}+ w_{10}P_{10}+w_{11}P_{11}+(wc_{01}+wc_{10})\cdot\min(\sqrt{P_{00} P_{11}},\sqrt{P_{10} P_{01}}).
\end{equation}

We observe that the separable states maximizing the witness satisfy $\sqrt{P_{00} P_{11}}<\sqrt{P_{10} P_{01}}$. Moreover, from eqs.~\eqref{observable} and \eqref{average} one easily sees that $(wc_{01}+wc_{10})=8 \alpha_1 \alpha_2 e^{-\alpha_1^2-\alpha_2^2}$ is non-negative. To simplify the expression we thus relax the separable bound using $\min(\sqrt{P_{00} P_{11}},\sqrt{P_{10} P_{01}})\leq \sqrt{P_{00} P_{11}}$. Computing all the coefficients $w$ from Eqs.~\eqref{observable} and \eqref{average} we obtain
\begin{align}\label{eq:wppt_SM}
\begin{split}
\tr(\rho_\mathrm{qubit}^{\mathrm{sep}} \hat{W}_\mathrm{qubit}) \leq w_\mathrm{ppt} &= \left(2 e^{-\alpha_1^2}-1\right) \left(2 e^{-\alpha_2^2}-1\right) P_{00} \\
  &\quad + 8 \alpha_1 \alpha_2 e^{-\alpha_1^2-\alpha_2^2} \sqrt{P_{00} P_{11}} \\
  &\quad + \left(2\alpha_1^2 e^{-\alpha_1^2} -1\right) \left(2\alpha_2^2 e^{-\alpha_2^2} -1\right) P_{11} \\
  &\quad + \left(2 \alpha_1^2 e^{-\alpha_1^2} -1\right) \left(2 e^{-\alpha_2^2}-1\right) P_{10} \\
  &\quad + \left(2 e^{-\alpha_1^2}-1\right) \left(2\alpha_2^2 e^{-\alpha_2^2} -1\right) P_{01}.
\end{split}
\end{align}
where here $\alpha_i\in \mathbb{R^+}$ denote the displacement amplitudes. If the quantity $V=(\langle{\hat{W}_\mathrm{qubit}}\rangle - w_\mathrm{ppt})$ is positive, we can conclude that the measured state is entangled under the condition that it is a 2-qubit state. \newline

To elaborate on the robustness of our witness with respect to losses on the state, let us consider the pure state 
\begin{equation}
\ket{\Psi^+}=\frac{1}{\sqrt{2}}(\ket{01}+\ket{10}).
\end{equation}
After traveling through a lossy channel with transmission $\eta$, this state becomes 
\begin{equation}
\rho_{\eta}=(1-\eta)\ketbra{00}{00}+\eta\ketbra{\Psi^+}{\Psi^+}.
\end{equation}
For such a state we have
\begin{equation}
V=8 \alpha_1 \alpha_2 e^{-\alpha_1^2-\alpha_2^2} \frac{\eta}{2}
\end{equation}
which is positive for all amplitude of the displacements. This means that $\hat{W}$ has the ability to detect entanglement for arbitrary loss on the state $\ket{\Psi^+}$. This statement can easily be generalized to every state of the form $\ket{\psi}=(\ket{01}+e^{i\phi}\ket{10})/\sqrt{2}$ by selecting the displacement parameters $\alpha_i \in \mathbb{C}$ in Eq.~\eqref{observable} accordingly.
In the next section we extend this witness to the case of an arbitrary Hilbert space. 

\section{Separable bound outside the qubit space}
\label{sec:app_bound2}
\subsection{Bound on $w_\text{ppt}$ for displacement with fluctuating amplitudes} 
In practice, the amplitude of the displacement might fluctuate during the experiment. Averaging the amplitudes for short times, we have access to a range of fluctuations $\alpha_1$ $\in$ $\mathcal{I}_1$, $\alpha_2$ $\in$ $\mathcal{I}_2$ with $\mathcal{I}_1,\mathcal{I}_2 \subset \mathbb{R^+}$. We thus simply consider 
\begin{align}\label{thebound}
\begin{split}
w_\mathrm{ppt} &\leq \max_{\alpha_1 \in \mathcal{I}_1,\alpha_2 \in \mathcal{I}_2}\Bigg(\left(2 e^{-\alpha_1^2}-1\right) \left(2 e^{-\alpha_2^2}-1\right) P_{00} \\
  &\qquad + 8 \alpha_1 \alpha_2 e^{-\alpha_1^2-\alpha_2^2} \sqrt{P_{00} P_{11}} \\
  &\qquad + \left(2 \alpha_1^2 e^{-\alpha_1^2} -1\right) \left(2 \alpha_2^2 e^{-\alpha_2^2} -1\right) P_{11} \\
  &\qquad + \left(2 \alpha_1^2 e^{-\alpha_1^2} -1\right) \left(2 e^{-\alpha_2^2}-1\right) P_{10} \\
  &\qquad + \left(2 e^{-\alpha_1^2}-1\right) \left(2 \alpha_2^2 e^{-\alpha_2^2} -1\right) P_{01} \Bigg).
\end{split}
\end{align}
In the case where we do not restrict ourselves to qubits, we replace $P_{00}$ and $P_{11}$ by $P_\mathrm{c,c}$ and $P_\mathrm{nc,nc}$. Furthermore we replace $P_{10}$ and $P_{01}$ by $(P_\mathrm{c,nc}-p_1^\ast)$ and $(P_\mathrm{nc,c}-p_2^\ast)$ if the maximization in Eq. ~\eqref{thebound} gives negative terms in front of $P_{10}$ and $P_{01}$ where $p_1^\ast>p(n_1\geq 2)$ is a bound on the probability to get strictly more than one photon in mode 1 and similarly for $p_2^\ast$. If the maximization in Eq.~\eqref{thebound} gives positive terms in front of $P_{10}$ and $P_{01}$, we replace them by $P_\mathrm{c,nc}$ and $P_\mathrm{nc,c}$, respectively. We thus get
\begin{equation}
w_\mathrm{ppt}\leq C_1 P_\mathrm{nc,nc}+C_2 \sqrt{P_\mathrm{nc,nc} P_\mathrm{c,c}}+C_3 P_\mathrm{c,c}+
\max \big(C_4 (P_\mathrm{c,nc}-p_1^\ast), C_4 P_\mathrm{c,nc} \big)+
\max \big(C_5 (P_\mathrm{nc,c}-p_2^\ast), C_5 P_\mathrm{nc,c} \big),
\end{equation}
where $C_1 ,$ $C_2 ,$ $C_3 ,$ $C_4 ,$ and $C_5 $ come from the maximization in Eq.~\eqref{thebound}.

\subsection{Bound on $\langle{\hat{W}}\rangle$ for separable state outside the qubit space} 
We consider the observable $\hat{W}$ represented by
\begin{equation}
\hat{W} =
\begin{pmatrix} 
A & B \\
B^{\dagger} & C
\end{pmatrix}
\end{equation}
and a state $\rho$ of the form
\begin{equation}
\rho =
\begin{pmatrix} 
a & b \\
b^{\dagger} & c
\end{pmatrix},
\end{equation}
where $a$ and $A$ live in the two-qubit space $\mathcal{H}_Q$ spanned by $\{\ket{0},\ket{1}\}$ and $c$ and $C$ live in a Hilbert space $\mathcal{H}_C$. We define $P_2$ as the trace of $c$ so that the trace of $a$ is equal to $(1-P_2)$. 
 The parts are now arranged such that
\begin{equation}
\label{trace}
\tr(\rho \hat{W})=\tr(a A)+\tr(b^{\dagger} B+b B^{\dagger} )+\tr(c C).
\end{equation}
We are interested in particular in the maximum expectation value of $\hat{W}$ a separable state can achieve. We treated the part which belongs to the qubit space above. We thus only have to find tight upper bounds on the quantities $\tr(b^{\dagger} B+b B^{\dagger})$ and $\tr(c C)$ that any state $\rho$ satisfies. \newline

Let us first deal with the term $\tr(c C)$, we have 
\begin{equation}
\tr (c\, C)= \tr \left(
\begin{pmatrix}
0 &0\\
0& c
\end{pmatrix} 
\hat W \right) \leq P_2 ||\hat W||_1 = P_2,
\end{equation}
where $||\hat W||_1$ is the maximal eigenvalue of $\hat W$, bounded by one by definition. This settles the issue. \newline

Next, consider the term $\tr( b B^\dag)$. We start with the singular value decomposition of $b$ 
\begin{equation}\label{eq:SVD}
    \begin{pmatrix} 
    0&b\\
    0&0
    \end{pmatrix}  =  
    \begin{pmatrix} 
    0&U DV^\dag\\
    0&0
    \end{pmatrix}
    = \sum_{i=1}^{L} d_i \ket{a_i}\! \bra{c_i},  
\end{equation}
where $\{\ket{a_i}\}$ and $\{\ket{c_i}\}$ are sets of orthonormal vectors on $\mathcal{H}_Q$ and $\mathcal{H}_C$ respectively, the singlular values $d_i$ are nonnegative real numbers, and $L=\text{min}\big( \text{dim}(\mathcal{H}_Q),\text{dim}(\mathcal{H}_C)\big)=4$ in our case. Using Eq.~\eqref{eq:SVD} we can bound
\begin{align}\label{eq:bB}
|\tr(b B^{\dagger})| &= \sum_{i=1}^L d_i| \bra{c_i}B^\dag \ket{a_i} |
 \leq \left(\sum_{i=1}^L d_i \right) b_\mathrm{max},
  \end{align}
where $b_\mathrm{max}$ is the maximal singular value of $B^\dag$, or equivalently of $B$, it can be easily obtained from the definition of the witness in Eq.~\eqref{average}, see later. It remains to upper-bound the term $(\sum_{i=1}^L d_i)$. \newline

To do so define a set of orthonormal states $\ket{\psi^i} \in \mathcal{H}_Q\oplus \mathcal{H}_C$ as 
\begin{equation}
    \ket{\psi^i} = \cos(\theta) \ket{a_i} + \sin(\theta) \ket{c_i}
\end{equation}
for some parameter $\theta$. As $\sum_{i=1}^L \ket{\psi^i} \! \bra{\psi^i}\leq \mathds{1}$ we obtain the following inequality
\begin{equation}
\label{overlap}
1\geq \sum_{i=1}^L\bra{\psi^i}\rho\ket{\psi^i}=\cos^2(\theta) \left(\sum_{i=1}^L  \bra{\chi_A^i} a
\ket{\chi_A^i}\right)+
\sin^2(\theta) \left(\sum_{i=1}^L \bra{\chi_C^i}c\ket{\chi_C^i}\right)
+2 \cos(\theta)\sin(\theta)\sum_{i=1}^L  d_i.
\end{equation}
Finally, by rearranging the terms we get the desired bound 
\begin{align}\nonumber
    \sum_{i=1}^L  d_i &\leq \frac{1 -\cos^2(\theta) \left(\sum_i \bra{\chi_A^i} a
\ket{\chi_A^i}\right) -\sin^2(\theta) \left(\sum_i \bra{\chi_C^i}c\ket{\chi_C^i}\right)}{2 \cos(\theta)\sin(\theta) }\\
&\leq \frac{1 -\cos^2(\theta) \tr (a) -\sin^2(\theta) \tr(c)}{2 \cos(\theta)\sin(\theta) } = \frac{1 -\cos^2(\theta) (1-P_2) -\sin^2(\theta) (P_2)}{2 \cos(\theta)\sin(\theta) }
\end{align}
for any value of $\theta$. Minimizing the right hand side with respect to $\theta$ yields the final bound
\begin{equation}
\sum_{i=1}^L d_i\leq \sqrt{P_2(1-P_2)}.\\
\end{equation}

We thus have access to the two following inequalities in order to bound $w_\mathrm{ppt},$
\begin{align}
\tr(c C) &\leq P_2, \\
\tr(b^{\dagger} B+ b B^{\dagger}) &\leq 2\sqrt{P_2(1-P_2)}|b_\mathrm{max}|,
\end{align}
where $P_2$ is the probability to have strictly more than one photon in at least one mode. For our witness, the phase randomization kills all the terms in $B$ except two, namely the photon-number preserving terms $2\sqrt{2}\alpha_1\alpha_2^3 \ketbra{11}{02}$ and $2\sqrt{2}\alpha_1^3\alpha_2\ketbra{11}{20}$. We thus simply have
\begin{equation}
b_\mathrm{max}=2 \alpha_1 \alpha_2 e^{-\alpha_1^2-\alpha_2^2} \sqrt{2 \alpha_1^4+2 \alpha_2^4}
\end{equation} 
for which we consider the maximum value in $\mathcal{I}_1$ and $\mathcal{I}_2$
\begin{equation}
\beta = \max_{\alpha_1 \in \mathcal{I}_1,\alpha_2 \in \mathcal{I}_2}\Big(b_\mathrm{max}\Big).
\end{equation}
The last part is to find a bound on $P_2$ one can measure in the experiment.
One has the probabilities
\begin{align}
P_2 &= p(n_1\leq 1  \cap n_2 > 1)+p(n_1>1  \cap n_2 \leq 1)+p(n_1> 1  \cap n_2 > 1), \\
p_1^\ast &\geq p(n_1 >1) = p(n_1>1  \cap n_2 \leq 1)  + p(n_1> 1  \cap n_2 > 1),\\
p_2^\ast &\geq p(n_2 >1) = p(n_1\leq 1  \cap n_2 > 1) +  p(n_1> 1  \cap n_2 > 1),
\end{align}
where $n_i$ denotes the number of photons in mode $i$. If we sum the two last quantities, we end up with an upper bound on $P_2$
\begin{align}
p_1^\ast+p_2^\ast &\geq 2p(n_1> 1  \cap n_2 > 1)+p(n_1\leq 1  \cap n_2 > 1)+p(n_1>1  \cap n_2 \leq 1) \nonumber \\
&= P_2+p(n_1> 1  \cap n_2 > 1) \nonumber \\
&\geq P_2.
\end{align}
In practice $p_i^\ast$ can be obtained by measuring the probability of coincidence after a 50/50 beam splitter. 
 All together with the first part, this implies that if a state $\rho$ obeys
\begin{align}\label{wpptmax}
\begin{split}
\tr(\rho \hat{W}) &> C_1 P_\mathrm{nc,nc}+ C_2 \sqrt{P_\mathrm{nc,nc} P_\mathrm{c,c}} + C_3 P_\mathrm{c,c} + \max \big(C_4 (P_\mathrm{c,nc}-p_1^\ast), C_4 P_\mathrm{c,nc} \big) \\
&\quad + \max \big(C_5 (P_\mathrm{nc,c}-p_2^\ast), C_5 P_\mathrm{nc,c} \big) + p_1^\ast+p_2^\ast+2\beta \sqrt{(p_1^\ast+p_2^\ast)(1-(p_1^\ast+p_2^\ast))}
\end{split}
\end{align}
then $\rho$ is entangled.

\section{Non unit detection efficiency}
\label{sec:app_deteff}
We considered so far that the measurement are realized with non-photon-number-resolving detectors preceded by displacement operations in phase space. As explained in the main text, we assign the outcome $+1$ to a no-detection and $-1$ to a conclusive detection event. Given a state $\rho$ in the mode 1 corresponding to the bosonic operators $a_1$ and $a_1^{\dagger}$, the probability to get a outcome $+1$ ($P_\mathrm{nc}$) using a displacement with argument $\alpha_1$ is given by
\begin{equation}
P_\mathrm{nc}=\tr(D^{\dagger}(\alpha_1)\ketbra{0}{0}D(\alpha_1)\rho).
\end{equation}
In the case where the detector has a finite efficiency, we can model the detector inefficiency with a beam splitter having a transmission $\eta=\cos{\varphi}^2,$ that is
\begin{equation}
P_\mathrm{nc}=\tr(D^{\dagger}(\alpha_1) U^{\dagger}\ketbra{\bar{0}}{\bar{0}} U D(\alpha_1)\rho)
\end{equation}
with $U = e^{\varphi(a^{\dagger}c-c^{\dagger}a)},$ the auxiliary mode described
by $c$ and $c^{\dagger}$, being initially empty. The state $\ket{\bar{0}}$ corresponds to the projection onto the vacuum for both modes. Commuting the beam splitter and displacement operation leads to
\begin{equation}
P_\mathrm{nc}=\tr(U^{\dagger}D^{\dagger}(\alpha_1\sqrt{\eta})\ketbra{\bar{0}}{\bar{0}} D(\alpha_1\sqrt{\eta})U\rho).
\end{equation}
This means that we can model the detection inefficiency as loss operating on the state that is measured if the amplitude of the displacement operation is changed accordingly. Hence, the fact that we consider detectors with unit efficiencies is still a valid description of our measurement apparatus where we do not need any assumptions on our state nor on the efficiency of our detectors.

\section{Finite statistic analysis}
\label{sec:app_finitestats}
All quantities that are measured are frequencies. We make the i.i.d hypothesis. For $N$ clicks on the heralded detectors, $n_a$ clicks on Alice detectors (and no clicks on Bob detector), $n_b$ clicks on Bob detectors (and no clicks on Alice detectors) and $n_d$ double clicks, we take the following estimators $\overline{P_\mathrm{c,nc}}=\frac{n_a}{N},$ $\overline{P_\mathrm{nc,c}}=\frac{n_b}{N},$ $\overline{P_\mathrm{c,c}}=\frac{n_d}{N}$ and $\overline{P_\mathrm{nc,nc}}=\left(1-\frac{n_a+n_b+n_d}{N}\right)$. We proceed in the same way for the estimator of $p_1^\ast$ and $p_2^\ast$. The corresponding standard deviations are
\begin{align}
\sigma_{c,c} &= \frac{\sqrt{\overline{P_\mathrm{c,c}}(1-\overline{P_\mathrm{c,c}})}}{\sqrt{N}}, 		& \sigma_{c,nc} &= \frac{\sqrt{\overline{P_\mathrm{c,nc}}(1-\overline{P_\mathrm{c,nc}})}}{\sqrt{N}},	& \sigma_{p_1^\ast} &= \frac{\sqrt{\overline{p_1^\ast}(1-\overline{p_1^\ast})}}{\sqrt{N}}, \nonumber\\
\sigma_{nc,nc} &= \frac{\sqrt{\overline{P_\mathrm{nc,nc}}(1-\overline{P_\mathrm{nc,nc}})}}{\sqrt{N}}, 	& \sigma_{nc,c} &= \frac{\sqrt{\overline{P_\mathrm{nc,c}}(1-\overline{P_\mathrm{nc,c}})}}{\sqrt{N}}, 	& \sigma_{p_2^\ast} &= \frac{\sqrt{\overline{p_2^\ast}(1-\overline{p_2^\ast})}}{\sqrt{N}}.\nonumber
\end{align} 
When the dependence of the witness with respect to the probabilities is linear, then the standard deviations can be added straightforwardly. Let us focus on the terms $\sqrt{P_\mathrm{nc,nc} P_\mathrm{c,c}}$. One cannot find an unbiased estimator for this term but one can bound it by a linear quantity
\begin{equation}
\sqrt{P_\mathrm{nc,nc} P_\mathrm{c,c}}\leq \frac{P_\mathrm{nc,nc}\overline{P_\mathrm{c,c}}+P_\mathrm{c,c}\overline{P_\mathrm{nc,nc}}}{2\sqrt{\overline{P_\mathrm{c,c}}~\overline{P_\mathrm{nc,nc}}}}.
\end{equation}
The same holds for the term $\sqrt{(p_1^\ast+p_2^\ast)~(1-(p_1^\ast+p_2^\ast))}$ which we bound according to
\begin{equation}
\label{bprob}
\sqrt{(p_1^\ast+p_2^\ast)~(1-(p_1^\ast+p_2^\ast))} 
\leq \frac{(p_1^\ast+p_2^\ast) - 2 (p_1^\ast+p_2^\ast) (\overline{ p_1^\ast+p_2^\ast}) + (\overline{ p_1^\ast+p_2^\ast})}{2 \sqrt{(\overline{ p_1^\ast+p_2^\ast})~(1- (\overline{ p_1^\ast+p_2^\ast}))}}.
\end{equation}
Note that Ineq.~\eqref{bprob} holds for $(p_1^\ast+p_2^\ast)\leq \frac{1}{2}$. The parts are now arranged such that an upper bound to the separable bound can be estimated according to
\begin{align}
\begin{split}
\overline{w}_\mathrm{ppt}^\mathrm{max} &= C_1 \overline{P_\mathrm{nc,nc}}+C_2\frac{\overline{P_\mathrm{nc,nc}}~\overline{P_\mathrm{c,c}}+\overline{P_\mathrm{c,c}}~\overline{P_\mathrm{nc,nc}}}{2\sqrt{\overline{P_\mathrm{c,c}}~\overline{P_\mathrm{nc,nc}}}}+ C_3 \overline{P_\mathrm{c,c}} + \max \big(C_4(\overline{P_\mathrm{c,nc}}-\overline{p_1^\ast}), C_4\overline{ P_\mathrm{c,nc}}\big) \\
&\quad + \max \big(C_5(\overline{P_\mathrm{nc,c}}-\overline{p_2^\ast}), C_5\overline{ P_\mathrm{nc,c}}\big)  + p_1^\ast+p_2^\ast+2 \beta \frac{(\overline{p_1^\ast+p_2^\ast}) - 2 (\overline{p_1^\ast+p_2^\ast})  (\overline{ p_1^\ast+p_2^\ast}) + (\overline{ p_1^\ast+p_2^\ast})}{2 \sqrt{(\overline{ p_1^\ast+p_2^\ast})~(1- (\overline{ p_1^\ast+p_2^\ast}))}}.
\end{split}
\end{align}
With the assumption of independent measurements for each probability, the standard deviation of this estimator is upper bounded by
\begin{align}
\begin{split}
\sigma_\mathrm{ppt}^\mathrm{max} &= C_1 \sigma_{nc,nc}+C_2 \frac{\sigma_{nc,nc}~\overline{P_\mathrm{c,c}}+\sigma_{c,c}~\overline{P_\mathrm{nc,nc}}}{2\sqrt{\overline{P_\mathrm{c,c}}~\overline{P_\mathrm{nc,nc}}}}+C_3 \sigma_{c,c}+|C_4|(\sigma_{c,nc}+\sigma_{p_1^\ast})+|C_5|(\sigma_{nc,c}+\sigma_{p_2^\ast}) \\
  &\quad+ \sigma_{p_1^\ast}+\sigma_{p_2^\ast}+2\beta\frac{(\sigma_{p_1^\ast}+\sigma_{p_2^\ast}) - 2 (\sigma_{p_1^\ast}+\sigma_{p_2^\ast}) (\overline{ p_1^\ast+p_2^\ast})+(\overline{ p_1^\ast+p_2^\ast})}{2 \sqrt{(\overline{ p_1^\ast+p_2^\ast})~(1- (\overline{ p_1^\ast+p_2^\ast}))}}.
\end{split}
\end{align}
In the same way, one can define an estimator for $\langle{\hat{W}}\rangle$
\begin{equation}
\overline{w}_\rho^\mathrm{exp} = \overline{P_\mathrm{c,c}}+\overline{P_\mathrm{nc,nc}}-\overline{P_\mathrm{c,nc}}-\overline{P_\mathrm{nc,c}}
\end{equation}
and the standard deviation of this estimator
\begin{equation}
\sigma_\rho^\mathrm{exp} = \sigma_{c,nc}+\sigma_{nc,c}+\sigma_{c,c}+\sigma_{nc,nc}.
\end{equation} 
We then say that we observe a violation of our witness by $k$ standard deviations if
\begin{equation}
\overline{w}_\mathrm{ppt}^\mathrm{max} - \overline{w}_\rho^\mathrm{exp} = k (\sigma_\mathrm{ppt}^\mathrm{max} + \sigma_\rho^\mathrm{exp}).
\end{equation}

\section{Phase locking requirement}
\label{sec:app_phaselocking}
\begin{figure}[b]
\capstart
\includegraphics[width = 7cm]{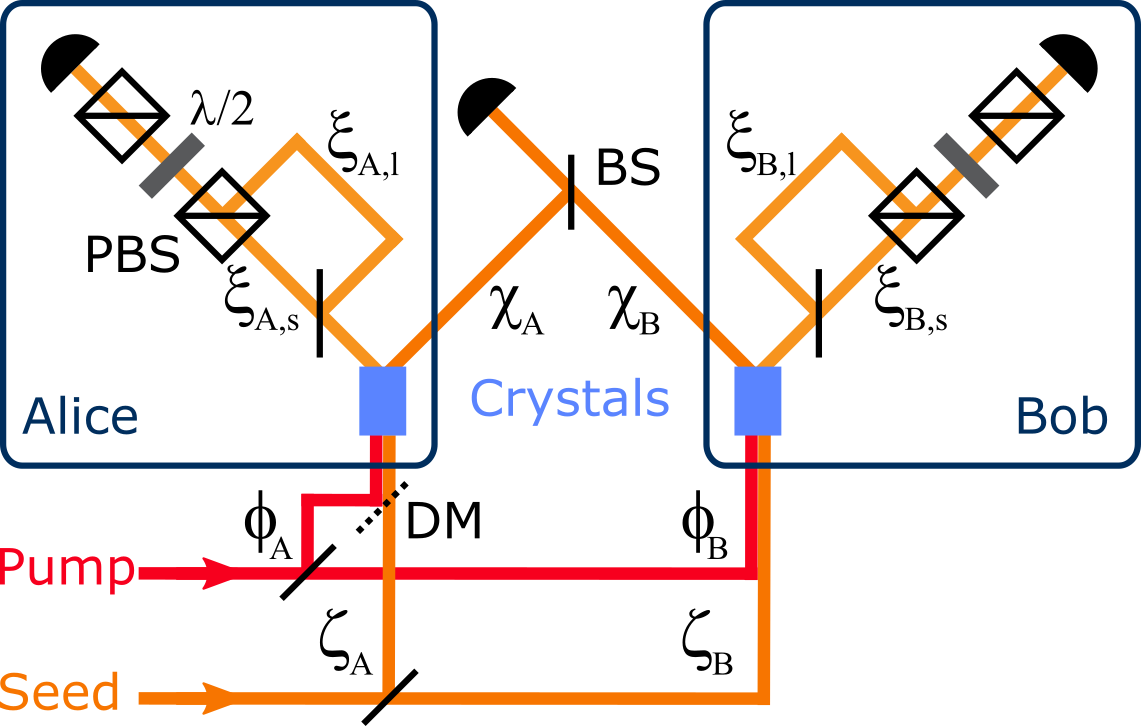}
\caption{\label{fig:Phases} Schematic of the experimental setup with the relevant phases. PBS, polarizing beam splitter; BS, beam splitter; DM, dichroic mirror; $\lambda /2$, half-wave plate.}
\end{figure}
In order to detect entanglement in a state of the form $\ket{\psi}=(\ket{01}+e^{i\phi}\ket{10})/\sqrt{2}$, our witness makes use of displacement operations on both modes (see Eq.~\eqref{observable}).
For such a measurement, the joint coincidence probabilities $P_{ij}$ will depend on optical phases acquired in the experimental setup. In the following, we will calculate the expected joint probability $P_{00}$ of having a no-click event in both modes for our experimental setup which will give rise to a phase locking requirement if we want to be able to detect entanglement. The calculation for joint probabilities different than $P_{00}$ would lead to the same requirement.
\begin{align}\label{eq:P00}
P_{00} &= \left|\bra{00}_{AB} D(\alpha_1)D(\alpha_2) \ket{\psi}_{AB}\right|^2 \nonumber \\
&= \left|\bra{00}_{AB} D(\alpha_1)D(\alpha_2)\frac{1}{\sqrt{2}}\left(e^{i(\phi_A+\chi_A+\xi_{A,l})}\ket{10}_{AB} + e^{i(\phi_B+\chi_B+\xi_{B,l})}\ket{01}_{AB}\right)\right|^2 \nonumber \\
&= \frac{1}{2}\left|e^{i(\phi_A+\chi_A+\xi_{A,l})}\bra{00}_{AB}D(\alpha_1)\ket{1}_A\ket{\alpha_2}_B + e^{i(\phi_B+\chi_B+\xi_{B,l})}\bra{00}_{AB}D(\alpha_2)\ket{\alpha_1}_A\ket{1}_B\right|^2,
\end{align}
where $D(\alpha_{1(2)})$ denotes the displacement operator acting on mode $A(B)$ with displacement parameters $\alpha_{1(2)} = |\alpha_{1(2)}| e^{i(\phi_{A(B)}-\zeta_{A(B)}+\xi_{{A(B)},s})}$. 
\begin{itemize}
\item $\phi_{A(B)}$ is the phase of the pump before the crystal, 
\item $\zeta_{A(B)}$ the phase of the seed laser before the crystal, 
\item $\chi_{A(B)}$ the phase picked up from the crystal to the central station,
\item $\xi_{A(B),l(s)}$ the phase from the crystal to the detector through the long (short) arm of the AMZI 
\end{itemize}
on Alice's (Bob's) side as shown in Fig.~\ref{fig:Phases}. This leads us to 
\begin{align}\label{eq:P00_2}
P_{00} &= \frac{1}{2} \left| e^{i(\phi_A+\chi_A+\xi_{A,l})}e^{-|\alpha_2|^2/2} \braket{-\alpha_1}{1}_A + e^{i(\phi_B+\chi_B+\xi_{B,l})}e^{-|\alpha_1|^2/2}\braket{-\alpha_2}{1}_B \right|^2 \nonumber \\
&= \frac{1}{2}\left|e^{i(\phi_A+\chi_A+\xi_{A,l})}e^{-(|\alpha_1|^2+|\alpha_2|^2)/2}(-\alpha_1^\ast) + e^{i(\phi_B+\chi_B+\xi_{B,l})}e^{-(|\alpha_1|^2+|\alpha_2|^2)/2}(-\alpha_2^\ast)\right|^2 \nonumber \\
&= \frac{1}{2}e^{-|\alpha_1|^2-|\alpha_2|^2}\left|e^{i(\phi_A+\chi_A+\xi_{A,l})}|\alpha_1|e^{-i(\phi_A-\zeta_A+\xi_{A,s})} + e^{i(\phi_B+\chi_B+\xi_{B,l})}|\alpha_2| e^{-i(\phi_B-\zeta_B+\xi_{B,s})}\right|^2
\end{align}
and setting $|\alpha|=|\alpha_1|=|\alpha_2|$ for simplicity yields
\begin{equation}\label{eq:P00_3}
P_{00} = \frac{1}{2}|\alpha|^2 e^{-2|\alpha|^2}\left|e^{i(\zeta_A+\chi_A+\xi_{A,l}-\xi_{A,s})} + e^{i(\zeta_B+\chi_B+\xi_{B,l}-\xi_{B,s})}\right|^2.
\end{equation}
We note that the phases of the pump before the crystals $\phi_A$ and $\phi_B$ cancel out. In order to keep $P_{00}$ constant, we therefore require
\begin{equation}\label{eq:phase}
\zeta_A+\chi_A+\xi_{A,l}-\xi_{A,s} = \zeta_B+\chi_B+\xi_{B,l}-\xi_{B,s}+\mathrm{const.}
\end{equation}
This can be achieved by locking the central interferometer according to
\begin{equation}\label{eq:phase1}
(\zeta_A+\chi_A) \mod{2\pi} = \zeta_B+\chi_B+\pi
\end{equation}
and the two asymmetric Mach-Zehnder interferometers (AMZI) for the displacement-based measurement such that
\begin{equation}\label{eq:phase2}
(\xi_{A,l}+\xi_{B,s}) \mod{2\pi} = \xi_{A,s}+\xi_{B,l}+\pi.
\end{equation}
\section{Experimental setup: phase locking of the measurement interferometers}
\label{sec:app_experiment}
\begin{figure}[b]
\capstart
\includegraphics[width = 17.2cm]{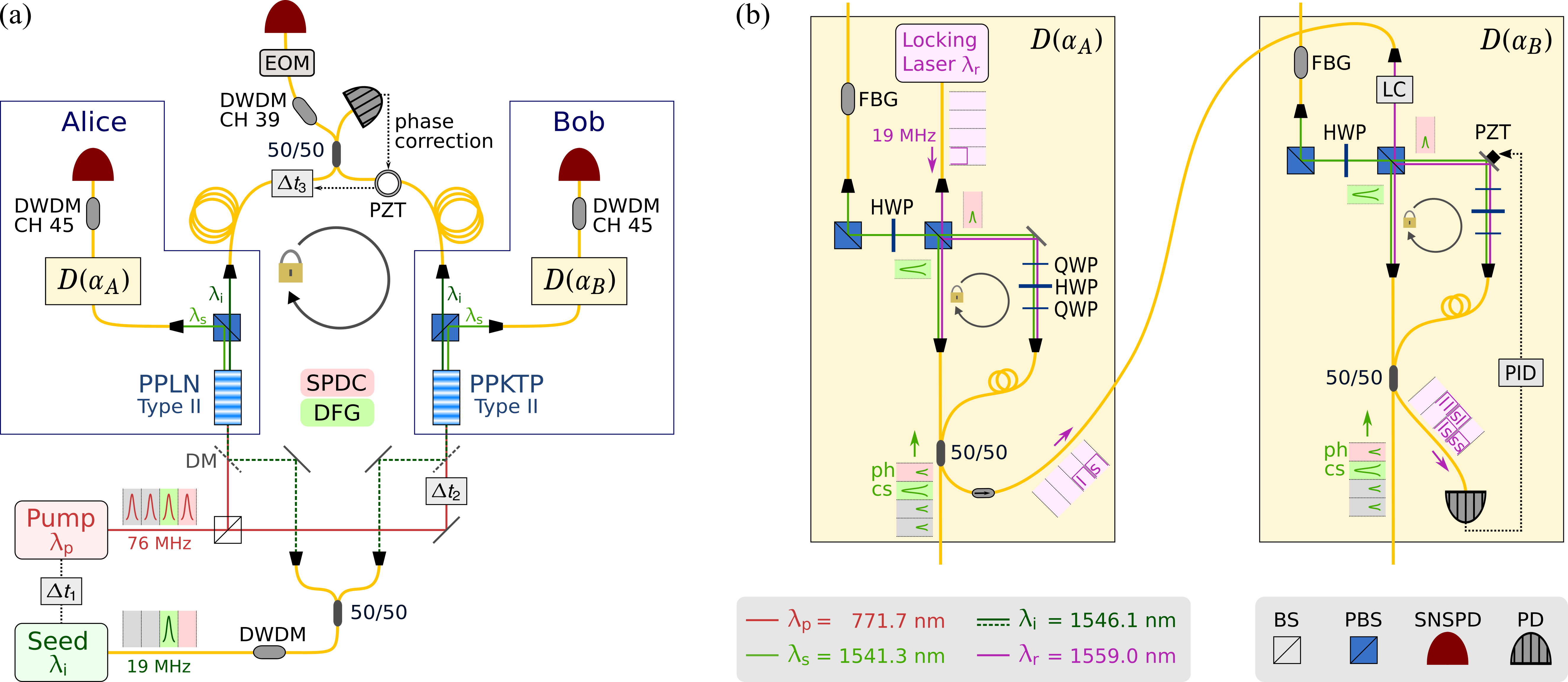}
\caption{\label{fig:SetupDetailed} (a) Simplified schematic of the experimental setup for the heralded distribution and certification of single-photon path entanglement. (b) Detailed schematic of the implementation for the displacement-based measurement using two asymmetric Mach-Zehnder interferometers (AMZI) with path difference corresponding to \SI{13}{ns}. The phase locking technique makes use of a pulsed DFB laser at $\lambda_r=\SI{1559.0}{nm}$ with a repetition rate of \SI{19}{MHz} and pulse duration of about \SI{10}{ns}. HWP, half-wave plate; QWP, quarter-wave plate.}
\end{figure}
A schematic overview of the experimental implementation is presented in Fig.~\hyperref[fig:SetupDetailed]{\ref{fig:SetupDetailed}(a)}.
To fulfill the crucial phase stability requirement derived in Sec.~\ref{sec:app_phaselocking}, we need to phase lock the central interferometer as described in the main text, and specific to our implementation of the displacement-based measurement, the phase difference between the AMZIs shown in Fig.~\hyperref[fig:SetupDetailed]{\ref{fig:SetupDetailed}(b)}. \par
We first note that in the common paths of the coherent state and the single-photon state, their short temporal separation of \SI{13}{ns} intrinsically guarantees the phase stability since the phase changes occurring over this time scale are negligible. 
However, the phase has to be actively stabilized in the AMZIs which temporally bring the single-photon and coherent states to coincidence. 
We therefore inject distributed feedback (DFB) laser pulses at $\lambda_r=\SI{1559.0}{nm}$ of about \SI{10}{ns} duration at a repetition rate of \SI{19}{MHz} traveling in the reverse direction to the signal, first through Alice's AMZI, then through Bob's, as schematically shown in Fig.~\hyperref[fig:SetupDetailed]{\ref{fig:SetupDetailed}(b)}. The resulting averaged signal shows \SI{50}{\%} visibility interference fringes as a function of the phase difference between the long-short and short-long paths in the measurement AMZIs, which is kept at a constant set-point by controlling a piezo-actuated mirror in the long arm of Bob's AMZI. \par 
The phase difference of the displacement fields between Alice and Bob is scanned and set with a liquid crystal (LC; Thorlabs LCC1111T-C) in the locking laser path before Bob's AMZI. The LC is aligned with the subsequent polarizing beam splitter (PBS) and therefore the signal in the long arm of Bob's AMZI is selectively retarded with respect to the signal in the short arm. In this way, the LC allows us to induce an additional relative phase between the displacement fields on Alice's and Bob's side over a range of $\pi$. In order to suppress unwanted reflections of the locking laser leaking to the detectors, we use fiber Bragg gratings (FBG) rejecting light at $\lambda_r$. \par 

\section{Spectral overlap and temporal alignment}
\label{sec:app_spectrum}
\begin{figure}[htb]
\capstart
\includegraphics[width = 17.2cm]{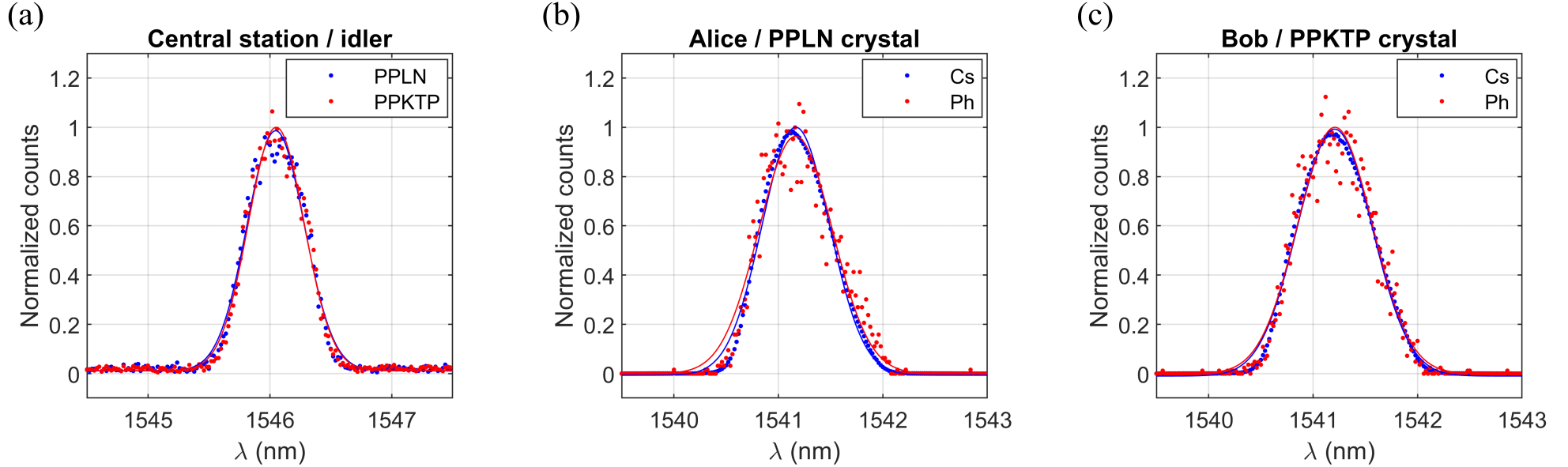}
\caption{\label{fig:Spectrum} Normalized measured spectra and Gaussian fit of (a) idler photons from independent sources before the heralding detector, (b) coherent state (Cs) and heralded signal photon (Ph) before Alice's detector and (c) before Bob's detector. All spectra are measured with a tunable grating filter with a FWHM of \SI{0.2}{nm} inserted before the corresponding superconducting nanowire single-photon detector.}
\end{figure}

To ensure high-purity heralded signal photons, we spectrally filter the heralding idler photons~\cite{McMillan2013} emitted from the PPLN crystal on Alice's side and from the PPKTP crystal on Bob's side after the 50/50 beam splitter by using a dense wavelength division multiplexer (DWDM) with a \SI{100}{GHz} passband at ITU channel 39 ($\lambda=\SI{1546.12}{nm}$).
The expected HOM visiblity due to finite spectral overlap between the idler photons (see Fig.~\hyperref[fig:Spectrum]{\ref{fig:Spectrum}(a)}) assuming Gaussian spectral distribution (see Eq.~5.14 in \cite{Mosley2007}) amounts to \SI{99.9}{\%}. \par
We also spectrally filter before Alice's and Bob's detectors with two DWDMs at channel 45 ($\lambda=\SI{1541.35}{nm}$) and achieve a spectral overlap between the single-photon and coherent states~\cite{Bruno2014} used for the displacement operation of more than \SI{99}{\%}, as shown in Figs.~\hyperref[fig:Spectrum]{\ref{fig:Spectrum}(b-c)}. \par

The fine temporal alignment of the central interferometer is achieved in the following way. First, the seed laser is replaced by a low coherence white light source and the motorized fiber delay line $\Delta t_3$ (see  Fig.~\hyperref[fig:SetupDetailed]{\ref{fig:SetupDetailed}(a)}) is set such that the observed interference visibility at the central station is maximized. Second, the white light source is exchanged by a cw laser at the signal wavelength $\lambda_s$ such that together with the pulsed pump laser, a coherent state at the idler wavelength $\lambda_i$ is created via difference frequency generation (DFG) in both crystals. The delay $\Delta t_2$ is then adjusted such that again the observed interference visibility at the central station is maximized. Third, the pulsed seed laser is put back in its place and the electronic delay $\Delta t_1$ is set such that pump and seed pulses overlap. This procedure ensures temporal indistinguishability of the idler photons from the two independent sources at the central station. 

\section{Results}
\label{sec:app_results}
The measured joint probabilities in the $\alpha$- and $z$-basis as well as the local probability of having more than one photon locally $p_i^\ast$ are given in Table~\ref{tab:ResultsProb}. In the case of $l=\SI{42}{m}$ of optical fiber inserted in each arm of the central interferometer, we measured lower probabilities for $P_\text{c,nc}$ and $P_\text{nc,c}$ in the $z$-basis compared to the case of $l=\SI{1.0}{km}$. These probabilities are direct measures of the signal photon transmissions. The difference can be explained by the realignment of the measurement AMZIs between the two experimental runs leading to an increase of the transmission on the signal photons in the case of $l=\SI{1.0}{km}$. \par 

The results of the entanglement witness are given in Table~\ref{tab:ResultsWitness_SM}. We detect photons at the central station at a heralding rate $\nu_h$ with the indicated signal-to-noise ration (SNR). 
The values for $w_\rho^\mathrm{exp}$ are computed from the joint probabilities in the $\alpha$-basis. We further calculate the values for the PPT bound $w_\mathrm{ppt}$ according to Eq.~\ref{eq:wppt_SM}, the bound including fluctuations of the displacement amplitudes $\widetilde{w}_\mathrm{ppt}$ according to the RHS of Ineq.~\ref{thebound} and the bound with additional contributions from outside the qubit space $w_\mathrm{ppt}^\mathrm{max}$ according to the RHS of Ineq.~\ref{wpptmax}. \par
For $l=\SI{42}{m}$, we measured at two different displacement amplitudes. The theoretical values for the amplitudes of the displacement parameters leading to the largest violation of our witness are $\alpha_1 = \alpha_2=1/\sqrt{2}\approx 0.71$, however, the witness is more robust to experimental fluctuations of the displacement parameter amplitudes if $\partial^2 w_\mathrm{ppt}/\partial\alpha_1\partial\alpha_2=0$, which in our case holds true for $\alpha_1 = \alpha_2 \approx 0.83$. This can be seen by comparing the difference between $\widetilde{w}_\mathrm{ppt}$ and $w_\mathrm{ppt}$ for the two $\alpha$-settings. \par 

We observe lower SNR for larger $\alpha$ as well as for longer fiber. We suspect this elevated noise background to be backward-scattered and forward-reflected Raman light that leaks through the DWDM before the heralding detector. One solution to this problem might be additional spectral filtering, however, leading to lower transmission on the heralding photons. 

\begin{table}[htb]
\capstart
\centering
\begin{ruledtabular}
\begin{tabular}{c c c c l l l l l l} 
$l$ 	& $\alpha_1$ 		& $\alpha_2$ 		& Basis 	& $P_\mathrm{nc,nc}$ 	& $P_\mathrm{nc,c}$ 	& $P_\mathrm{c,nc}$ 	& $P_\mathrm{c,c}$ 	& $p_1^\ast$ 				& $p_2^\ast$\\	
\midrule
\multirow{4}{*}{\SI{42}{m}} & \multirow{2}{*}{$0.720^{+0.014}_{-0.010}$} 	& \multirow{2}{*}{$0.710^{+0.008}_{-0.008}$} 	& $z$ 	& 0.96834(4) 	& 0.01431(3) 	& 0.01735(3) 	& 0.0000044(5) 	& \multirow{4}{*}{\num{2.5(3)E-6}} 	& \multirow{4}{*}{\num{5.1(4)E-6}} \\	\cmidrule{4-8}
	&				&				& $\alpha$  	& 0.3604(2) 			& 0.2305(2) 			& 0.2407(2) 			& 0.1684(2) 			& 					& \\			
\cmidrule{2-8}
	& \multirow{2}{*}{$0.804^{+0.010}_{-0.009}$} 	& \multirow{2}{*}{$0.819^{+0.003}_{-0.004}$} 	& $z$ 		& 0.96935(5) 		& 0.01515(3) 		& 0.01550(3) 		& 0.0000052(6)		& 					 	& \\	\cmidrule{4-8}
	&					&					& $\alpha$  	& 0.2715(2) 			& 0.2504(2) 			& 0.2393(2) 			& 0.2388(2) 			& 				& \\
\cmidrule{1-10}
\multirow{2}{*}{\SI{1.0}{km}} 	& \multirow{2}{*}{$0.819^{+0.005}_{-0.007}$} 	& \multirow{2}{*}{$0.837^{+0.006}_{-0.007}$} 	& $z$ 		& 0.96142(5) 		& 0.01881(4) 		& 0.01977(4) 		& 0.0000059(6)	& \multirow{2}{*}{\num{3.2(4)E-6}} 	& \multirow{2}{*}{\num{1.25(8)E-5}} \\  \cmidrule{4-8}
	&					&					& $\alpha$ 	& 0.2575(2) 			& 0.2504(2) 			& 0.2370(2) 			& 0.2552(2) 			& 						&
\end{tabular}
\end{ruledtabular}
\caption{\label{tab:ResultsProb} Measured joint probabilities in the $z$- and $\alpha$-basis for different fiber lengths $l$ in each arm of the central interferometer. The displacement parameter amplitudes $\alpha_i$ are the mean amplitudes for the $\alpha$-basis measurement with bounds on the minimum and maximum observed values during the \SI{3600}{s} of measurement. The probabilities of having more than one photon $p_1^\ast$ on Alice's side and $p_2^\ast$ on Bob's side are separately determined for each experimental run by measuring the heralded $g^2(0)$. The uncertainties on all probabilities are 1 standard deviations as calculated in Sec.~\ref{sec:app_finitestats}.}
\end{table}

\begin{table}[htb]
\capstart
\centering
\begin{ruledtabular}
\begin{tabular}{c c c c c c c c c c} 
$l$ 					& $\nu_h$ (kHz) & SNR 	& $\alpha_1$ 		& $\alpha_2$ 	& $w_\rho^\mathrm{exp}$ & $w_\mathrm{ppt}$ & $\widetilde{w}_\mathrm{ppt}$ & $w_\mathrm{ppt}^\mathrm{max}$ 	& $k$ \\
\midrule
\multirow{2}{*}{\SI{42}{m}} & 1.4 	& 18 			& $0.720^{+0.014}_{-0.010}$ & $0.710^{+0.008}_{-0.008}$ & 0.0576(8) 			& 0.0391(2) 		& 0.0451(2) 	& 0.0472(14) 		& 4.8 \\
\cmidrule{2-10}
 					& 1.4 	& 12 			& $0.804^{+0.010}_{-0.009}$ & $0.819^{+0.003}_{-0.004}$ & 0.0206(8) 			& 0.0039(2) 		& 0.0045(2) 	& 0.0071(16) 		& 5.6 \\ 
\cmidrule{1-10}
\SI{1.0}{km} 			& 1.6 	& 5.0 		& $0.819^{+0.005}_{-0.007}$ & $0.837^{+0.006}_{-0.007}$ & 0.0253(7) 			& 0.0031(2) 		& 0.0033(2) 	& 0.0071(22) 		& 6.2
\end{tabular}
\end{ruledtabular}
\caption{\label{tab:ResultsWitness_SM} Measured expectation value of the entanglement witness $w_\rho^\mathrm{exp}$ and calculated separable bound $w_\mathrm{ppt}$, including fluctuations of the displacement amplitudes $\widetilde{w}_\mathrm{ppt}$ and additional contributions from outside the qubit space $w_\mathrm{ppt}^\mathrm{max}$ for fibers of length $l$ inserted in each arm of the central interferometer at an observed heralding rate $\nu_h$. The indicated heralding rate includes noise that we observe with the indicated signal-to-noise ratio (SNR). The witness is violated by $k=(w_\rho^\mathrm{exp}-w_\mathrm{ppt}^\mathrm{max})/(\sigma_\mathrm{ppt}^\mathrm{max}+\sigma_\rho^\mathrm{exp})$ standard deviations.}
\end{table}

\end{document}